\documentclass[fleqn,usenatbib,useAMS]{mnras}

\usepackage[a4paper]{geometry}
\usepackage{newtxtext,newtxmath}

\usepackage[T1]{fontenc}
\usepackage{ae,aecompl}
\usepackage{hyperref}

\usepackage{graphicx}	
\usepackage{amsmath}	
\usepackage{amssymb}	
\usepackage{multicol}        
\usepackage{bm}		
\usepackage{pdflscape}	
\usepackage{fixltx2e}
\usepackage{rotating}
\usepackage{color}
\usepackage[update]{epstopdf}

\title[Radio Emission from Supermassive Hot Jupiters]{Supermassive Hot Jupiters Provide More Favourable Conditions for the Generation of Radio Emission via the Cyclotron Maser Instability - A Case Study Based on Tau Bootis b}

\author[C. Weber et al.]{
C.~Weber$^{1,2}$,\thanks{E-mail: cw@ufa.cas.cz, christof.weber@oeaw.ac.at} N.~V.~Erkaev,$^{3,4}$, V.A. Ivanov,$^{4}$, P.~Odert$^{1,5}$, J.-M.~Grie\ss meier$^{6,7}$,
\newauthor L. Fossati$^1$, H. Lammer$^{1}$, and H.~O.~Rucker$^8$
\\
$^{1}$Space Research Institute, Austrian Academy of Sciences, Schmiedlstr. 6, A-8042, Graz, Austria\\
$^{2}$Institute of Atmospheric Physics, Czech Academy of Sciences, Prague, Czech Republic (cw@ufa.cas.cz)\\
$^{3}$Institute of Computational Modelling SB RAS, 660036, Krasnoyarsk, Russian Federation\\
$^{4}$Siberian Federal University, Krasnoyarsk, Russian Federation\\
$^{5}$Institute of Physics/IGAM, University of Graz, Universit\"atsplatz 5, A-8010 Graz, Austria\\
$^{6}$LPC2E - Universit\'{e} d'Orl\'{e}ans/CNRS, France \\ 
$^{7}$Station de Radioastronomie de Nan\c{c}ay, Observatoire de Paris, PSL Research University, CNRS/INSU, USR 704 - Univ. Orl\'{e}ans, OSUC, \\ route de Souesmes, 18330 Nan\c{c}ay, France\\
$^{8}$Comm. Astron., Austrian Academy of Sciences, Graz, Austria
}

\date{Accepted XXX. Received YYY; in original form ZZZ}

\pubyear{2018}

\begin{document}
\label{firstpage}
\pagerange{\pageref{firstpage}--\pageref{lastpage}}
\maketitle

\begin{abstract}
We investigate under which conditions supermassive hot Jupiters can sustain source regions for radio emission, and whether this emission could propagate to an observer outside the system. We study Tau Bootis b-like planets (a supermassive hot Jupiter with 5.84 Jupiter masses and 1.06 Jupiter radii), but located at different orbital distances (between its actual orbit of 0.046 AU and 0.2 AU). Due to the strong gravity of such planets and efficient radiative cooling, the upper atmosphere is (almost) hydrostatic and the exobase remains very close to the planet, which makes it a good candidate for radio observations. We expect similar conditions as for Jupiter, i.e.~a region between the exobase and the magnetopause that is filled with a depleted plasma density compared with cases where the whole magnetosphere cavity is filled with hydrodynamically outward flowing ionospheric plasma. Thus, unlike classical hot Jupiters like the previously studied planets HD 209458b and HD 189733b, supermassive hot Jupiters should be in general better targets for radio observations.   
\end{abstract}




\section{Introduction}

It is known that the solar system planets emit low-frequency, coherent, polarized radio emission \citep[e.g.][]{Zarka1998,Farrell1999,Ergun2000,Treumann2006}, where electric fields parallel to the magnetic field accelerate electrons towards the planet into a region of increasing magnetic field. The electrons are partially reflected upwards and some are lost to the planetary atmosphere. The reflected electrons exhibit an unstable distribution, e.g.~a loss cone or a horseshoe distribution \citep[e.g.][]{Treumann2006}. The process leading to (exo)planetary radio emission from this electron distribution is the CMI (Cyclotron Maser Instability). It works only if the local electron plasma frequency is smaller than the local electron cyclotron frequency, $\omega_p < \omega_c$. In particular, it works most efficiently if $\omega_p \ll \omega_c$. Thus, the strongest radio emission comes from regions of low plasma density and strong magnetic field.

It is natural to assume that this process is also operating for planets outside the solar system. Various observation campaigns and theoretical studies investigated the possible radio emission from extrasolar planets \citep[see][and references therein]{Weber2017}. Here, we only highlight two recent theoretical studies by \citet{Griessmeier2017} and \citet{Lynch2018}. They extended the studies on detectability of exoplanetary radio emission by \citet{Lazio2004,Griessmeier2007b} and \citet{Griessmeier2011} to the exoplanet census at the time of their respective studies. \citet{Lynch2018} predict the planets V830 Tau b, BD$+$20 1790 b and GJ 876 b to generate observable levels of radio emission. \citet{Griessmeier2017} find e.g. Eps Eridani b and Tau Bootis b to be good candidates.

While close-in hot Jupiters have long been considered favourable candidates for observing radio emission, it has recently been pointed out by \citet{Weber2017,Weber2017b} and \citet{Yates2017,Yates2018} that close-in planets can have hydrodynamically expanded upper atmospheres and ionospheres due to the strong stellar XUV (X-ray and extreme ultraviolet) irradiation. These ionospheres can be very dense, leading to a high local plasma frequency. In some cases, this can lead to quenching of the planetary radio emission in the sense that either no emission can be generated, or that the emission is absorbed in the planetary ionosphere, which extends up to the magnetopause. For a planet of Jupiter's mass or less, one way to prevent this effect would be to look at planets at a larger distance from the star. This, however, weakens the interaction between the star and the planet and the energy input into the planetary magnetosphere, which is usually assumed to be essential for the generation of intense radio emission (a few alternative models do not rely on the proximity of the planet to the host star, see e.g.~\citet{Griessmeier2017} for a comparison). Another way to avoid planets with hydrodynamically extended atmospheres and ionospheres would be to study planets around stars with a weaker XUV flux, like the massive A5 star WASP-33 which hosts a hot Jupiter orbiting at a distance of 0.02 AU \citep[e.g.][]{Herrero2011}, or massive gas giants with several Jupiter-masses where the gravity may act against hydrodynamic expansion of their upper atmospheres.

Recently, \citet{Hallinan2015} detected aurorae from a brown dwarf. Since brown dwarfs are very massive their atmospheres should be very compact and they should not have an extended upper atmosphere and ionosphere, but rather a magnetosphere with vast regions of depleted plasma. Thus, they should exhibit source regions where the detected emission is produced and radio waves can freely propagate to an observer on e.g.~Earth. Such kind of objects belong to the so-called Ultra Cool Dwarfs and they form a bridge between hot Jupiters and stars \citep{Route2016b,Llama2018}. \citet{Hallinan2008} confirmed that these dwarf stars can have strong dynamos leading to magnetic fields strong enough to generate radio emission via the electron cyclotron maser instability \citep[e.g.][]{Treumann2006}. The authors observed an M-dwarf and an L-dwarf, detecting 100 \% circularly polarized and coherent radio emission requiring magnetic fields in the kG-range. These observations confirmed the CMI to be the dominant mechanism for the generation of radio emission in magnetospheres of Ultra Cool Dwarfs \citep{Hallinan2008}. Several other authors reported successful observations \citep{Berger2010,Williams2013,Route2016a,Route2016b,Williams2017}.

Since radio emission from Ultra Cool Dwarfs has been already detected and they bridge stars with hot Jupiters, it seems natural to investigate if supermassive hot Jupiters provide better conditions for the CMI than the classical Hot Jupiters studied in \citet{Weber2017}. Thus, in this work we are investigating the possible generation of radio emission from Tau Bootis b and the propagation of radio waves in the planetary vicinity. Since the first theoretical studies of exoplanetary radio emission, this planet has frequently been considered as one of the best candidates for radio observations \citep[e.g.~][]{Farrell1999, Griessmeier2005, Griessmeier2006b, Griessmeier2007c, Griessmeier2007b, Reiners2010, Griessmeier2011, Griessmeier2017}. It has been the target of a number of radio observational campaigns \citep[e.g.][]{Bastian2000,Farrell2003,Ryabov2004,Lazio2004,Shiratori2006,Winterhalter2006,Majid2006,Lazio2007,Stroe2012,Hallinan2012}, and more observations of this planet have recently been performed with LOFAR \citep[e.g.][]{Turner2017b}. 

Here, Tau Bootis b will serve as the archetype of a certain class of planets, i.e.~supermassive hot Jupiters. The planet has a minimum mass ($M \sin i$) of 4.13 Jupiter masses, with an estimated true mass of 5.84 Jupiter masses (\url{www.exoplanet.eu}, accessed 2018-06-01). There are currently 92 known exoplanets with more than 2 Jupiter masses and orbital distances of less than 0.1 AU. 

Twelve of these planets are located at distances $\leq$ 100 pc from Earth. Supermassive hot Jupiters like Tau Bootis b may constitute some of the most promising candidates for radio observations. First, they are likely to have strong magnetic fields. In fact, all magnetic field estimations indicate that massive, gaseous planets should have stronger magnetic fields than less massive Jupiter-like gas giants \citep{Griessmeier2004, Reiners2010}. This increases the chance to generate radio emission with frequencies above the Earth's ionospheric cutoff of 10 MHz and increases the power that can be received by wide-band radio observations. Secondly, their high masses (and thus larger gravity) lead to more compact upper atmospheres. 

For planets more massive than Jupiter, like Tau Bootis b, the planetary gravity keeps the atmosphere strongly bound. In some cases, this will lead to a hydrostatic rather than hydrodynamic upper atmosphere \citep[for example, WASP-18b, is one of these cases;][]{Fossati2018}, so that radio emission could be generated and may escape, a situation which is comparable with the known solar system conditions for e.g.~Earth. In other words, a more massive planet maintains hydrostatic conditions closer to its host star than less massive, Jupiter-like planets. This has an added benefit for observational campaigns: close-in planets are easier to study from an observer's point of view, as it is easier to cover one full planetary orbit. In fact, the planetary radio emission is assumed to vary with the planetary orbit, which is one of the ways to discriminate between stellar and planetary emission \citep[see e.g.][]{Griessmeier2005}. The variation is caused by the inhomogeneity of both the planetary magnetic field and the stellar wind (and its magnetic field). For this reason, the stellar rotation has also to be taken into account \citep{Fares2010}, which is particularly important if the stellar rotation and the planetary orbital motion are tidally synchronized \citep[which seems to be the case for Tau Bootis;][]{Donati2008}. Whether the gravity is indeed strong enough to prevent the quenching of planetary radio emission has to be studied on a case-by-case basis; in this follow-up study to \citet{Weber2017}, we investigate the case of Tau Bootis b.

Section \ref{sec:sec21} describes briefly the upper atmosphere model used for evaluating the electron density profiles shown in Section \ref{sec:sec3} and shows the planetary and stellar parameters. This section also describes the upper atmosphere structure of Tau Bootis b at different orbit locations. The estimation of the magnetopause standoff distance as well as the stellar wind parameters are shown in Section \ref{sec:sec22}. This section also shows a comparison of standoff distance and exobase level. Section \ref{sec:sec3} addresses the results for the plasma environment and the corresponding plasma and cyclotron frequencies. Section \ref{sec:sec5} shortly addresses the implications of our study for future radio observations of Tau Bootis b. Section \ref{sec:sec6} presents our conclusions.

%

\section{Model description}\label{sec:sec2}

\subsection{Upper Atmosphere Modelling}\label{sec:sec21}

In \citet{Weber2017}, we used upper atmosphere profiles from hydrodynamic models which include heating by stellar XUV radiation, ionization and dissociation of hydrogen. This is a valid approach for most hot Jupiters which have expanded and efficiently escaping atmospheres. 

In cases of extremely massive planets, an upper atmosphere in the hydrodynamic regime is hardly possible, and thus it is reasonable to apply equilibrium equations for the atmospheric parameters. We derive these equations from the hydrodynamic system in \citet{Erkaev2016} by neglecting terms with time derivatives and bulk velocities. The equations for pressure $P$ and temperature $T$ are as follows
\begin{eqnarray}
\frac{\partial P}{\partial R} =
-\rho\frac{\partial U}{\partial R} ,    \label{P1}
\end{eqnarray}
\begin{eqnarray}
 Q_{\rm{EUV}} - Q_{\rm{cool}} + \frac{\partial }{R^2\partial R}\left(R^2 \chi \frac{\partial T}{\partial R}\right) =0.                              \label{T}
\end{eqnarray}
Here, $R$ is the distance from the planet, $U$ is the gravitational potential,
$\chi$ is the thermal conductivity of the atmosphere \citep{Watson1981}, given by
 \begin{equation}
 \chi = 4.45\cdot 10^4 \left(\frac{T}{1000 \textnormal{K}}\right)^{0.7},
 \end{equation}
$Q_{\rm{EUV}}$ is the stellar EUV (extreme ultraviolet) volume heating rate
\begin{equation}
Q_{\rm{EUV}} = \eta \sigma_{\rm{EUV}}\left(n_{\rm H} + n_{\rm H_2}\right)\phi_{\rm{EUV}} ,
\end{equation}
$\eta$ is the ratio of the net local heating rate to the rate of the stellar radiative absorption in the planetary atmosphere \citep[assumed to be 15 \%, in agreement with][]{Shematovich2014} and $Q_{\rm{cool}}$ is the cooling rate. The latter consists of the Lyman-$\alpha$ ($Q_{\rm{L_\alpha}}$),
\begin{equation}
Q_{\rm{L_\alpha}} = 7.5\cdot 10^{-19} n_{\rm e} n_{\rm H} \exp(-118348 \textnormal{K}/T),
\end{equation}
the $\rm{H}$ collision ionization ($Q_{\rm{ci}}$),
Bremsstrahlung ($Q_{\rm{BS}}$) and $\rm{H}_{\rm{+}}$ radiative recombination ($Q_{\rm{rec}}$)
cooling processes, i.e.~$Q_{\rm{cool}} = Q_{\rm{L_\alpha}} + Q_{\rm{ci}} + Q_{\rm{BS}} + Q_{\rm{rec}}$. The quantity $\sigma_{\rm{EUV}}$ is the cross section of EUV absorption.

The cooling rates $Q_{\rm {ci}}$, $Q_{\rm{BS}}$ and $Q_{\rm{rec}}$ are taken from \citet{Glover2007}.
$\phi_{\rm{EUV}}$ is the function describing the EUV flux absorption in the atmosphere
\begin{equation}
\phi_{\rm EUV}=\frac{1}{2}\int_0^{\pi/2+\arccos(1/r)} J_{\rm EUV}(r,\theta)\sin(\theta)d\theta.
\end{equation}
Here, $J_{\rm EUV}(r,\theta)$ is a function of spherical coordinates that describes the spatial variation of the  EUV flux  due to the atmospheric absorption \citep{Erkaev2015}, $r$ corresponds to the radial distance from the planetary
center normalized to the planetary photospheric radius $R_{\rm 0}$ .

The mass density, $\rho$, and the pressure, $P$,
are related to the species densities as follows:
\begin{equation}
\rho=m_{\rm H}\left(n_{\rm H} + n_{\rm H^+}\right) + m_{\rm H_2}\left(n_{\rm H_2} + n_{\rm H_2^+}\right),   \label{rho}
\end{equation}
\begin{equation}
P=\left(n_{\rm H}+n_{\rm H^+}+n_{\rm H_2}+n_{\rm{H_2^+}}+n_{\rm e}\right)kT,    \label{P2}
\end{equation}
where $k$ is the  Boltzmann constant, and $m_{\rm H}$ and $m_{\rm H_2}$ are the
masses of the hydrogen atoms and molecules, respectively.

Steady-state densities of atomic and molecular hydrogen, ions and electrons satisfy the following algebraic equations:
\begin{eqnarray}
\nu_{\rm H} n_{\rm H} + \nu_{\rm Hcol} n_{\rm e} n_{\rm H}-
\alpha_{\rm H} n_{\rm e}n_{\rm H^+} =0,          \label{nH}
\end{eqnarray}
\begin{eqnarray}
-\nu_{\rm H_2} n_{\rm H_2} - \nu_{\rm dis} n_{\rm H_2}n +
\gamma_{\rm H} n n_{\rm H}^2 =0,                     \label{nH2}
\end{eqnarray}
\begin{eqnarray}
\nu_{\rm H_2} n_{\rm H_2} - \alpha_{\rm H_2} n_{\rm e}n_{\rm H_2^+}=0 .  \label{nH+}
\end{eqnarray}
The electron density is determined by the quasi-neutrality condition
\begin{eqnarray}
n_{\rm e}=n_{\rm H^+}+n_{\rm H_2^+}                              \label{ne}
\end{eqnarray}
and the total hydrogen number density is a sum of the partial densities
\begin{eqnarray}
n=n_{\rm H} + n_{\rm H^+} + n_{\rm H_2} + n_{\rm H_2^+}.      \label{ntot}
\end{eqnarray}
The term $\alpha_{\rm H}$ is the recombination rate related to the reaction \mbox{H$^{+}+e\rightarrow \rm{H}$} of $4\times 10^{-12} (300 \textnormal{K}/T)^{0.64}$
cm$^{3}$ s$^{-1}$, $\alpha_{\rm H_2}$ is the dissociation rate of \mbox{H$_2^+ $+$e\rightarrow$H + H}: $\alpha_{\rm H_2}$=$2.3\times 10^{-8} (300 \textnormal{K}/T)^{0.4}$
cm$^{3}$ s$^{-1}$,
$\nu_{\rm diss}$ is the thermal dissociation rate of \mbox{H$_2$ $ \rightarrow$ H + H}: 1.5 $\cdot$ 10$^{-9}$ $\exp(-49000 \textnormal{K}/T)$,
$\gamma_{\rm H}$ is the rate of the reaction \mbox{H + H $\rightarrow$ H$_2$}: $\gamma_{\rm H}$ = 8.0 $\cdot $ 10$^{-33}$ (300 \textnormal{K}/T)$^{0.6}$
\citep{Yelle2004}.

The term $\nu_{\rm H}$ is the
hydrogen ionization rate, and $\nu_{\rm H_2}$ is the ionization rate of molecular hydrogen \citep{Storey1995, MurrayClay2009},
\begin{eqnarray}
\nu_{\rm H} = 5.9 \cdot 10^{-8} \phi_{\rm EUV} {\rm s^{-1}} , \quad \nu_{\rm H_2} = 3.3\cdot 10^{-8} \phi_{\rm EUV} {\rm s^{-1}},
\end{eqnarray}
and  $\nu_{\rm Hcol}$  is the collisional ionization rate \citep{Black1981}, $\nu_{\rm Hcol}$ = 5.9$\cdot 10^{-11} T^{1/2} \exp(-157809 \textnormal{K}/T)$ .

It turns out that the $\rm{H}_2^+$ density is much smaller than that of other species.
This allows us to neglect $n_{\rm H_2^+}$ in the equations for the electron and total densities (Equations \ref{ne} and \ref{ntot}). With this assumption, we solve the Equations (\ref{nH}-\ref{ntot}) and express the total, molecular, ion and electron densities through one quantity - the atomic hydrogen density.
Substituting these expressions to (\ref{P2}) and (\ref{rho}), we solve finally two ordinary differential equations (\ref{P1}, \ref{T}).

Table \ref{tab:tab1} summarizes the Tau Bootis b system parameters. Table \ref{tab:tab1b} shows the XUV fluxes at orbital distances of 0.046, 0.1 and 0.2 AU from Tau Bootis as well as the temperature $T_0$ and pressure $P_0$ at the photosphere. 

The lower boundary is set at the planetary photospheric radius $R_{\rm 0}$. For all cases, and throughout this study, we consider that $R_{\rm 0}$ lies at an atmospheric pressure $P_{\rm 0}$ of about 100 mbar. This is justified by the fact that, for an H$_{\rm 2}$-dominated atmosphere considering a clear atmosphere and taking into account H$_{\rm 2}$ Rayleigh scattering, H$_{\rm 2}$-H$_{\rm 2}$ collisional-induced absorption, alkali lines, and solar-abundance molecular bands, the optical depth at visible wavelengths is unity at a pressure level of about 100 mbar \citep{Lammer2016,Fossati2017,Cubillos2015,Cubillos2017}. The atmospheric temperature $T_{\rm 0}$ at $R_{\rm 0}$ is assumed to be the equilibrium temperature at the planet's orbital location (see Table \ref{tab:tab1b}).
 
\begin{table}
	\centering
	\caption{Tau Bootis system parameters. The stellar and planetary parameters are from \url{http://exoplanet.eu/catalog/tau_boo_b/}, accessed on 01.06.2018.}
	\label{tab:tab1}
	\begin{tabular}{p{3 cm}p{2 cm}} 
		\hline
		Planet & \\
		\hline
		Orbital distance & 0.046 AU \\
		Distance from Earth & 15.6 pc \\
		Mass & $5.84 M_{\rm J}$ \\
		Radius $R_0$ & $1.06 R_{\rm J}$\\
		Dipole moment $\mathcal{M}$ & $0.76 \mathcal{M_{\rm J}}$$^\ast$\\
		\hline
		\hline
		Star & Tau Bootis\\
		\hline
		Spectral type & F7 V\\
		Mass & $1.3 M_{\rm{Sun}}$\\
		Radius & $1.331 R_{\rm{Sun}}$\\
		Age & 2.52 Gyr\\
		\hline
		\hline
		\begin{footnotesize}* \citep{Griessmeier2007c}\end{footnotesize}
	\end{tabular}
\end{table}

To infer the XUV flux of Tau Bootis, we have first derived the chromospheric emission at the core of the CaII H\&K lines, for which high-quality observations are available, and converted it to an XUV flux value. We have first computed the synthetic spectral energy distribution of the stellar photosphere employing the LLmodels stellar atmosphere code \citep{Shulyak2004} to which we have then added various levels of emission at the core of the CaII H\&K lines, as described in \citet{Fossati2015}, until we could fit the optical observations obtained with the ESPADONS spectropolarimeter available in the CFHT archive. The star Tau Bootis is located much closer than 100 pc, thus interstellar medium absorption does not significantly affect the CaII line core emission \citep{Fossati2017}. In this way, we obtained an integrated CaII chromospheric emission flux at a distance of 1 AU from the star of 0.05 W/m$^2$. We have converted this flux into a Ly$\alpha$ flux and then into an XUV flux using the scaling relations of \citet{Linsky2013,Linsky2014}, obtaining an XUV flux of 17.5 W/m$^2$ at the distance of the planet. Being based on two scaling relations, this value has an uncertainty of the order of 50\% \citep[e.g.][]{Fossati2015}.

\begin{table}
	\centering
	\caption{Stellar XUV radiation at different orbital separations for Tau Bootis. Planetary atmospheric temperature and pressure $R_{\rm 0}$ of a Tau Bootis b-like planet at different orbital distances.}
	\label{tab:tab1b}
	\begin{tabular}{p{2 cm}p{1 cm}p{1.5 cm}p{1.2 cm}} 
		\hline
		Orbital distance [AU] & XUV [W/m$^2$] & Temperature $T_0$ [K] & Pressure $P_0$ [mbar]\\
		\hline
		\hline
		0.046 & 17.53 & 1700 & 100 \\
		\hline
		0.1 & 3.71 & 1150 & 100 \\
		\hline
		0.2 & 0.93 & 810 &  100 \\
		\hline
		\hline
	\end{tabular}
\end{table}

\begin{figure*}
\begin{center}
\includegraphics[width=0.8\columnwidth]{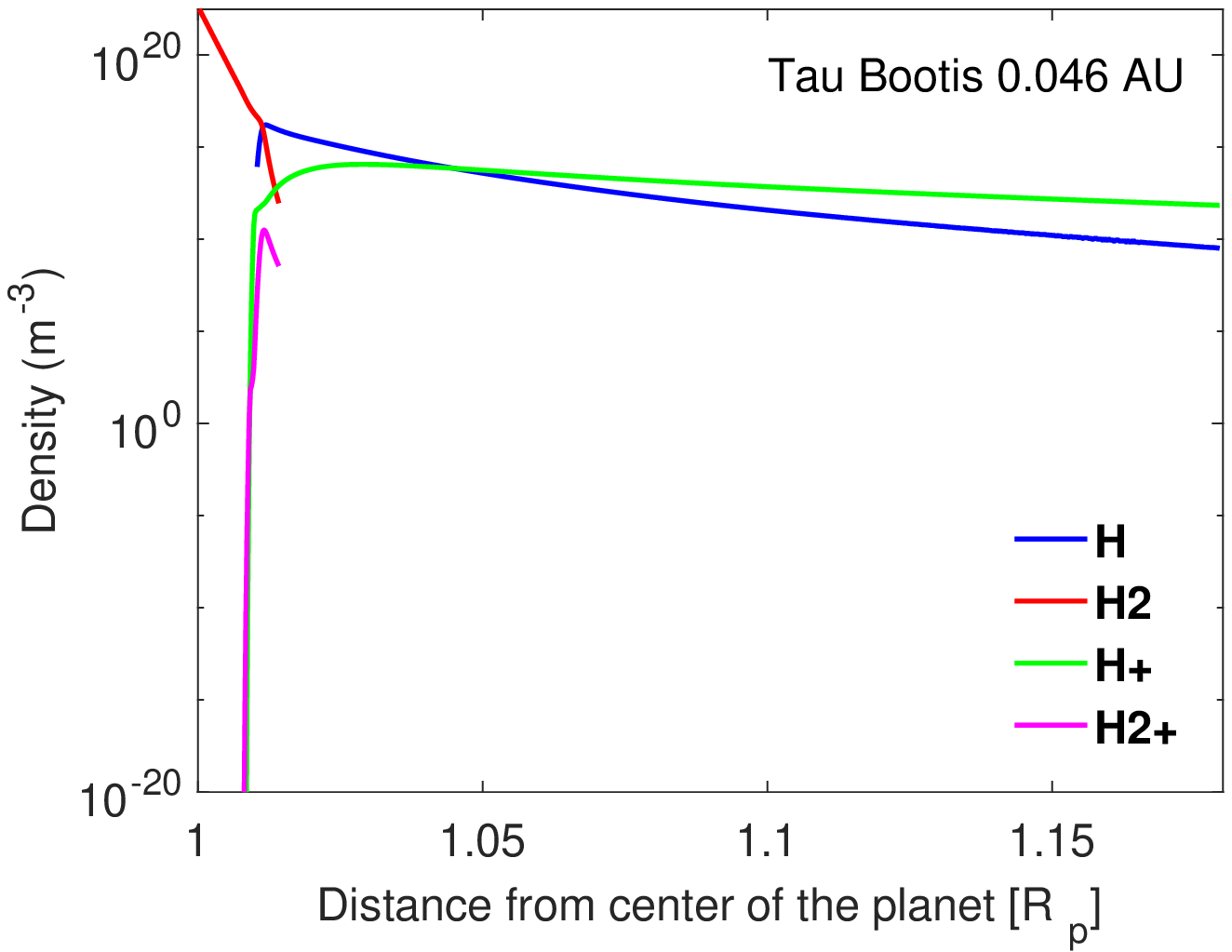}
\includegraphics[width=0.8\columnwidth]{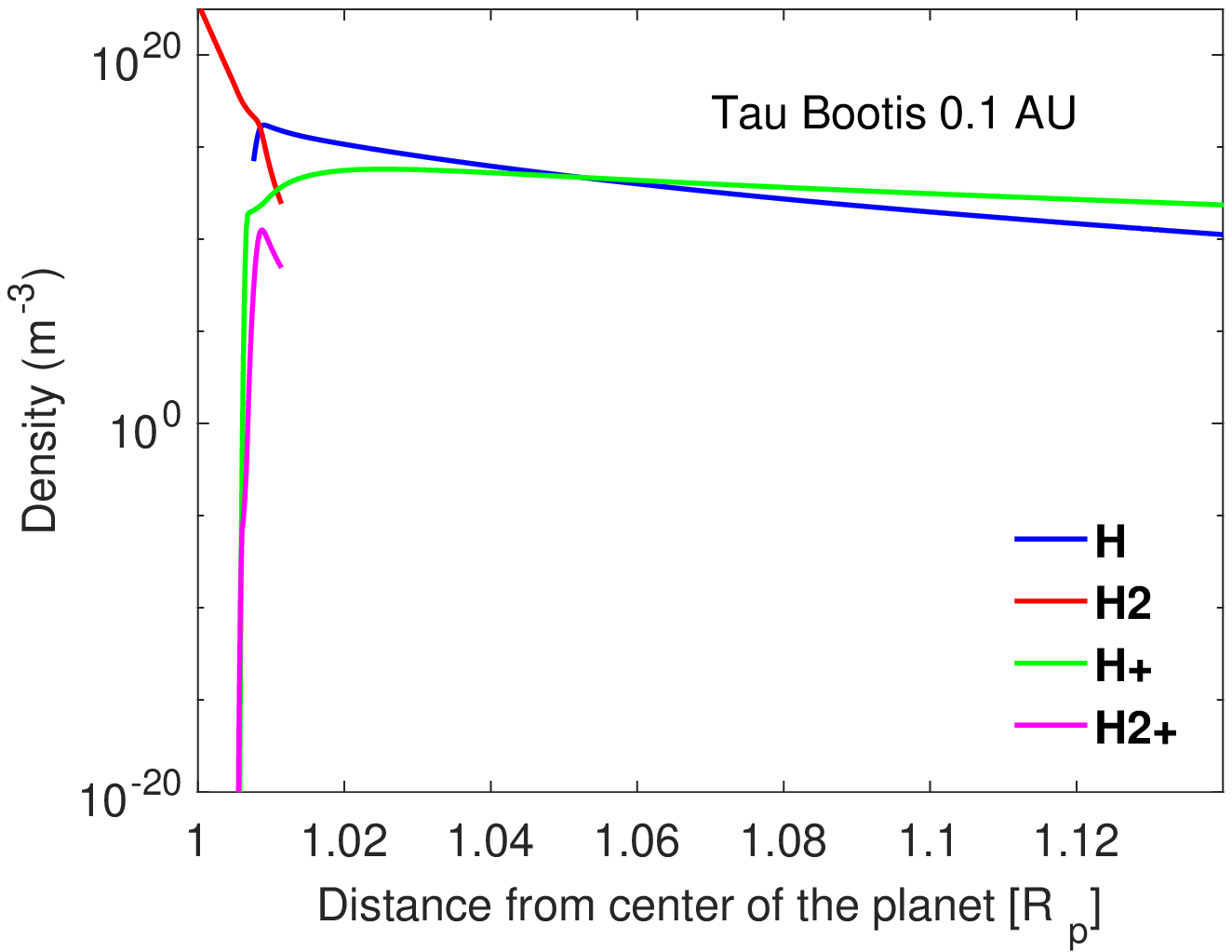}
\includegraphics[width=0.8\columnwidth]{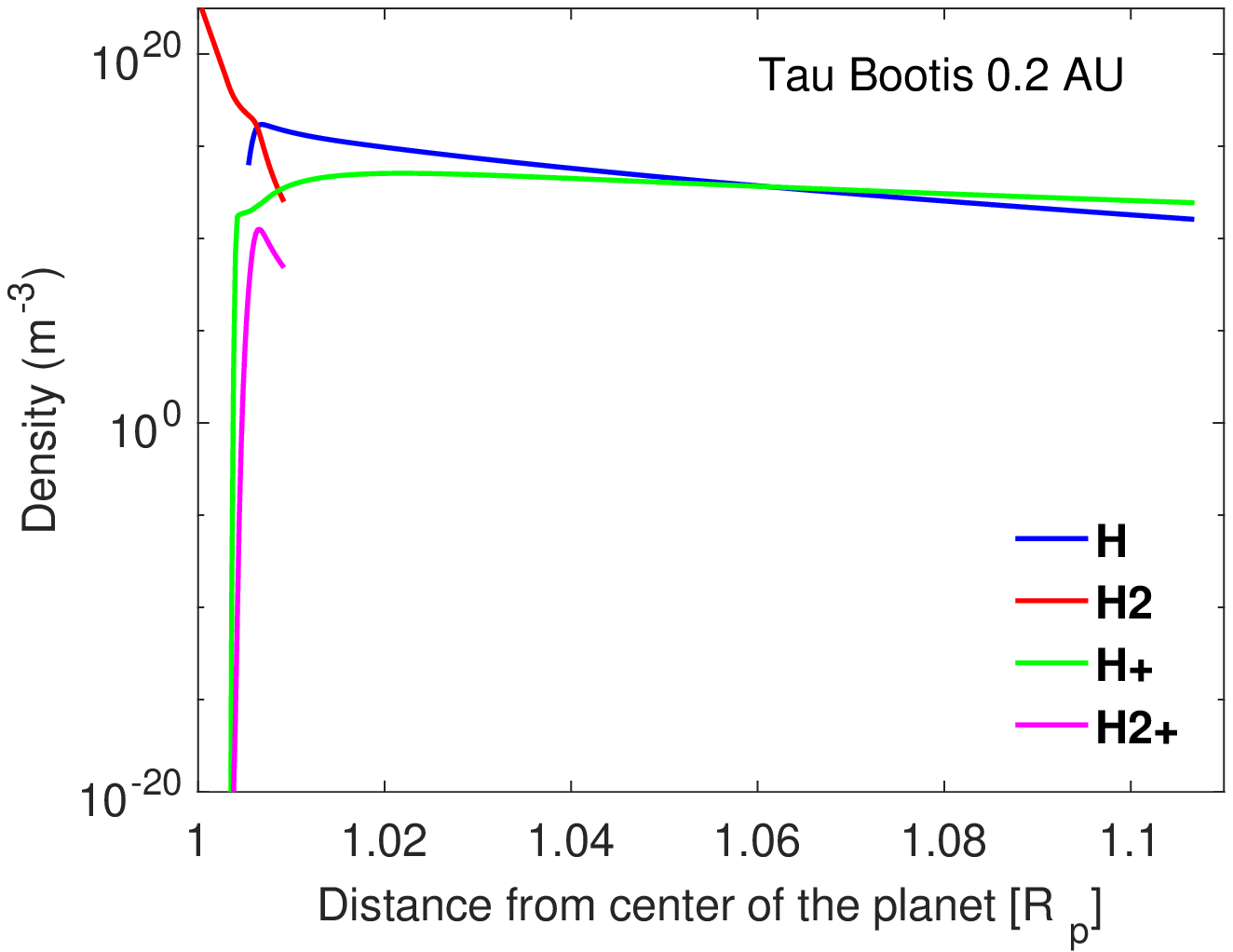}
\includegraphics[width=0.8\columnwidth]{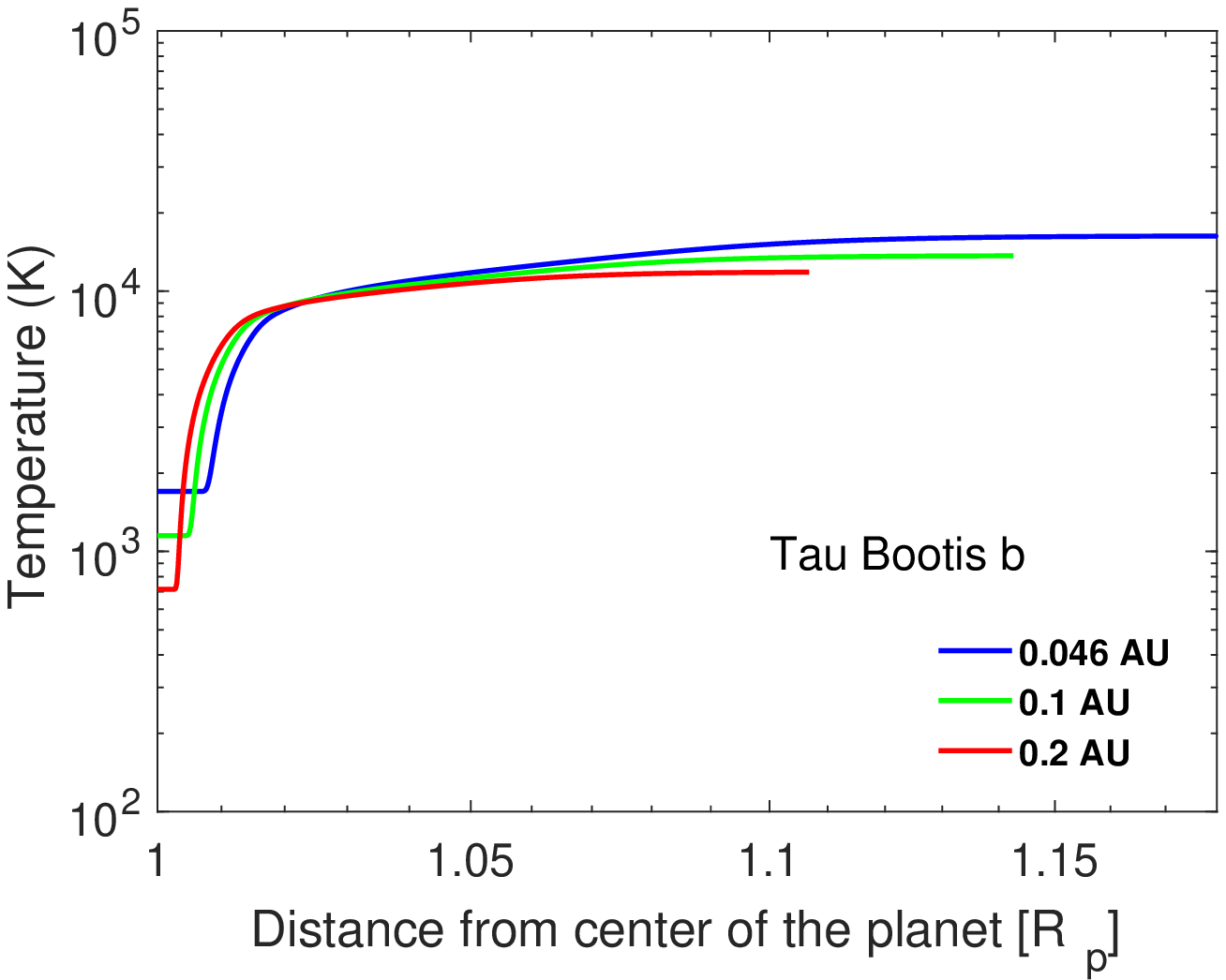}
\caption{Top panels and bottom-left panel: H, H$_{\rm 2}$, H$^+$ and H$_2^+$ densities at orbital distances of 0.046, 0.1 and 0.2 AU for Tau Bootis b around the F7V star Tau Bootis. Note that we only plot the regions up to the exobase level. Bottom-right panel: temperature profiles for 0.046, 0.1 and 0.2 AU.}
\label{fig:fig01b}
\end{center}
\end{figure*}

Figure \ref{fig:fig01b} shows the number density profiles of H, H$_2$, H$^+$ and H$_2^+$ (i.e., neutrals and plasma) at 0.046, 0.1 and 0.2 AU for Tau Bootis b (upper panels and lower left panel) and the temperature profile for the different orbital distances (right lower panel). The temperature peak for Tau Bootis b which would be needed for hydrodynamic escape is about $\sim 100000$ K, i.e. one order of magnitude higher than for the previously studied hot Jupiter HD 209458b. Such a high temperature is not realistic, because cooling processes can reduce it immediately. In case of a supermassive planet, the temperature peak is so high that the cooling processes make a significant contribution. Therefore the cooling processes reduce the temperature peak below the critical level needed for a hydrodynamic escape regime. When trying to run a hydrodynamic code with cooling for the case of a supermassive planet, there is no radial acceleration of the atmospheric particles because the temperature is not sufficiently high for such acceleration against the gravitational forces. We plan to investigate this effect in detail within a follow-up study. The atmosphere is not expanding hydrodynamically and very favourable conditions for the CMI can be expected. The strong gravity and the radiative cooling keep the atmosphere compact and we expect similar conditions as for e.g.~Jupiter, with large regions of depleted plasma between the exobase and the magnetopause.

By estimating the hydrogen scale height ($H=kT_{\rm{exo}}/m_{\rm H} g$, where $k$ is the Boltzmann constant, $T_{\rm{exo}}$ the temperature at the exobase, $m_{\rm H}$ the hydrogen mass and $g$ the gravitational acceleration) at the exobase level of Tau Bootis b and Jupiter we obtain 4443 km and 365 km, respectively. The distance between the exobase and the magnetopause for Jupiter is 40.9 $R_{\rm J}$ or 7.8 $\cdot 10^3$ scale heights and 3.6 $R_{\rm p}$ or about 60 scale heights for Tau Bootis b. One can expect that the exosphere density of Tau Bootis b decreases fast so that the plasmasphere between the exobase and possible magnetopause distances will be populated mainly with stellar wind plasma. Of course the stellar wind plasma at the orbit of Tau Bootis b is much higher than at 5.2 AU but this poses no problem for the propagation of possibly generated radio waves \citep{Griessmeier2007b}. Therefore, the upper atmosphere-magnetosphere configuration and the related plasmasphere are  more comparable with Jupiter's environment in the Solar System, making the production of radio waves by the CMI and propagation in that environment more likely compared to classical hot Jupiters like HD 209458b or HD 189733b.

\subsection{Comparison between exobase and magnetopause standoff distance}\label{sec:sec22}

The magnetopause standoff distances are calculated from \citep[e.g.~][]{Griessmeier2004, Khodachenko2012, Kislyakova2014}
\begin{equation}
	\label{eqn:magmom}
	R_{\rm s} = \left(\frac{\mathcal{M}^2\mu_0 f_0^2}{8\pi^2\rho_{\rm sw}v_{\rm sw}^2}\right)^{\frac{1}{6}}.
\end{equation}
Here, $\mu_0$ is the vacuum permeability and $f_0 = 1.22$ is a form factor for the magnetopause shape including the influence of a magnetodisk \citep{Khodachenko2012}. $v_{\rm{sw}}$ is the stellar wind velocity corrected for the orbital motion of the planet. $\rho_{\rm sw}$ is the stellar wind density and $\mathcal{M}$ is the magnetic moment. Table \ref{tab:swp} summarizes the stellar wind parameters from the model of \citet{Griessmeier2007c,Griessmeier2007b} for the different orbital distances.

\begin{figure}
	\includegraphics[width=\columnwidth]{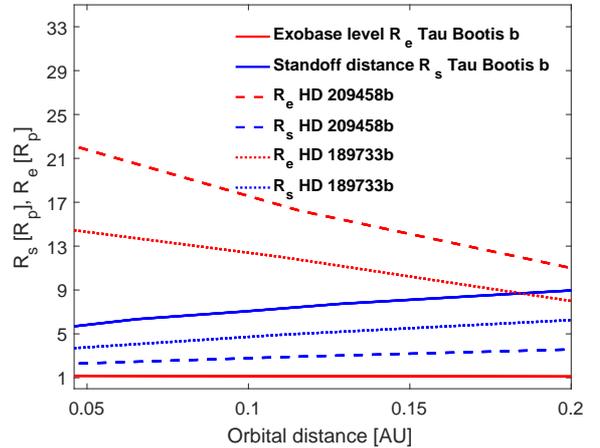}
    \caption{Exobase levels $R_{\rm e}$ compared to the magnetopause standoff distances $R_{\rm s}$ as a function of orbital distance for the magnetic moments as predicted by \citet{Griessmeier2007c} for Tau Bootis b and the planets studied in \citet{Weber2017}.}
    \label{fig:fig1tau}
\end{figure}

Figure \ref{fig:fig1tau} shows the exobase levels $R_{\rm e}$ compared to the magnetopause standoff distances $R_{\rm s}$ as a function of orbital distance for the magnetic moments as predicted by \citet{Griessmeier2007c} for Tau Bootis b and the planets studied in \citet{Weber2017}, i.e.~HD 209458b and HD 189733b. For Tau Bootis b, the exobase is very close to the planet (at 1.18, 1.14 and 1.11 $R_{\rm p}$ for 0.046, 0.1 and 0.2 AU, respectively) and the magnetospheric cavity for the CMI to work can be expected to be very large. For all orbital distances between 0.046 and 0.2 AU the exobase is smaller than the standoff distance. For HD 209458b and HD 189733b, assuming the same XUV flux as for Tau Bootis b, for each of the investigated orbits the exobase is larger than the magnetopause standoff distance.

\begin{table}
	\centering
	\caption{Stellar wind parameters (velocity $v_{\rm{sw}}$ and density $n_{\rm{sw}}$) of Tau Bootis at different orbital distances.}
	\label{tab:swp}
	\begin{tabular}{p{2 cm}|p{2 cm}|p{2 cm}} 
		Orbital distance [AU] & $v_{\rm{sw}}$ [km/s] & $n_{\rm{sw}}$ [m$^{-3}$] \\
		\hline
		\hline
		0.046 & 272 & $4.16 \cdot 10^{10}$ \\
		\hline
		0.1 & 340 & $6.04 \cdot 10^9$\\
		\hline
		0.2 & 408 & $1.22 \cdot 10^{9}$
	\end{tabular}
\end{table}  

\section{Plasma environment and corresponding frequencies}\label{sec:sec3}

For the cyclotron frequency, we test three different hypotheses for the planetary magnetic moment: (a) we use the magnetic field strengths predicted by \citet{Griessmeier2007c}, (b) we compare to the value of \citet{Reiners2010}, and (c) we also compare to results using the rotation-independent value of \citet{Griessmeier2011}. To calculate the corresponding cyclotron frequency $f_{\rm {c}}$ we use
\begin{equation}
	\label{eqn:freqbrelation1}
	f_{\rm {c}} = \frac{1}{2\pi}\frac{e\cdot B}{m_{\rm e}}.
\end{equation}
Here, $e$ is the electron charge, $B$ the magnetic field strength and $m_{\rm e}$ the electron mass. The relation between magnetic dipole moment and the maximum magnetic field strength at the pole is given by
\begin{equation}
	\label{eqn:maxind}
	B = \frac{\mu_0}{4\pi}\frac{\mathcal{M}}{R_{\rm p}^3},
\end{equation}
where $R_{\rm p}$ is the planetary radius, $\mu_0$ is the vacuum permeability and $\mathcal{M}$ is the magnetic moment. The plasma frequency is calculated via
\begin{equation}
	\label{eqn:fp1}
	f_{\rm p} = \frac{1}{2\pi}\sqrt{\frac{e^2n_{\rm e}}{m_{\rm e}\varepsilon_0}},
\end{equation}
with $n_{\rm e}$ the electron density and $\varepsilon_0$ the vacuum permittivity. For the electron density at Tau Bootis b, results from Section \ref{sec:sec21} are evaluated.

The plasma and cyclotron frequency are calculated to check whether the condition for the Cyclotron Maser Instability to work (i.e. $f_{\rm p} \ll f_{\rm {c}}$) is fulfilled \citep{Weber2017, Weber2017b}. 

\citet{Griessmeier2007c} predicted the magnetic moment of Tau Bootis b to be $0.76 \mathcal{M}_{\rm J}$, where $\mathcal{M}_{\rm J}$ is the Jovian magnetic moment. For the magnetic moment, we also considered the model by \citet{Reiners2010}, in which tidal locking has no influence. They predict a dipole magnetic field strength at the pole of 58 G. This corresponds to a magnetic moment of $1.25 \cdot 10^{28} \textnormal{Am}^2 \approx 7.5 \mathcal{M}_{\rm J}$. The value found in \citet{Griessmeier2011} corresponds to $\approx$ 20.3 G and a magnetic moment of $2.8 \mathcal{M}_{\rm J}$.

A recent study by \citet{Yadav2017} states that magnetic moments of up to 10 times stronger than those of Jupiter can be expected for hot Jupiters regardless of their age, provided that the energy of the stellar radiation is deposited in the planetary center. With the processes suggested here this is, however, not a realistic assumption. In the more realistic case of energy deposition in the planet's outer layers, the extra energy would reduce the thermal gradient within the planet or even invert it. This would reduce the convection in the planet. Rather than strengthening the planetary dynamo, this would weaken it, and could even shut it down altogether.

\begin{figure}
	\includegraphics[width=0.9\columnwidth]{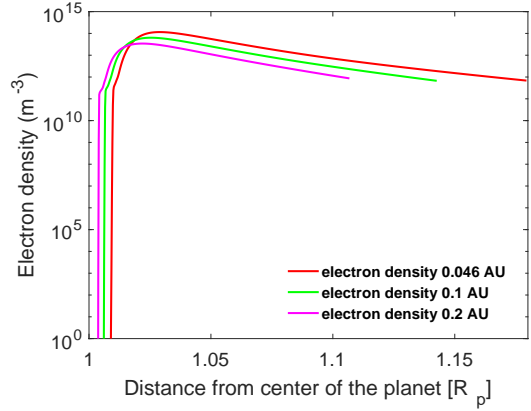}
    \caption{Electron density profiles of a Tau Bootis b-like planet at orbital distances of 0.046 (red), 0.1 (green) and 0.2 AU (magenta). The curves end at the exobase level.}
    \label{fig:fig2tau1}
\end{figure}

\begin{figure}
	\includegraphics[width=0.9\columnwidth]{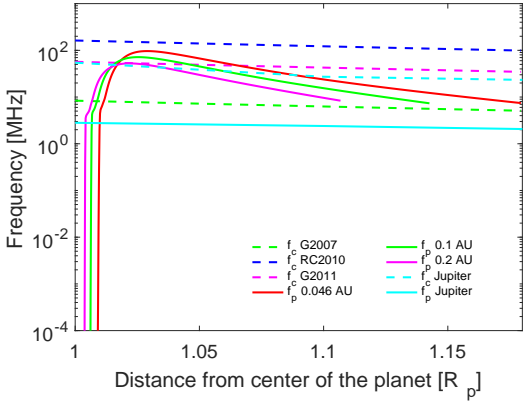}
    \caption{Comparison of cyclotron frequency to plasma frequency. The red, green and magenta lines correspond to the plasma frequency at 0.046, 0.1 and 0.2 AU, respectively, and stop at the exobase levels. The dashed green line shows the cyclotron frequency corresponding to a magnetic moment as predicted by \citet{Griessmeier2007c}. The dashed blue line shows the cyclotron frequency corresponding to a polar magnetic field strength of 58 G, as predicted by \citet{Reiners2010}. The dashed magenta line corresponds to the prediction by \citet{Griessmeier2011}. The solid and dashed cyan lines correspond to the plasma and cyclotron frequencies at Jupiter, respectively. The plotting range stops at the largest exobase level ($1.18 R_{\rm{p}}$ at 0.046 AU).}
    \label{fig:fig3tau1}
\end{figure}

Figure \ref{fig:fig2tau1} shows the electron density profiles of Tau Bootis b placed at 0.046 AU (red line), 0.1 AU (green line) and 0.2 AU (magenta line). Figure \ref{fig:fig3tau1} shows the cyclotron frequency (green, blue and magenta dashed lines) and the plasma frequency at different orbital distances (red, green and magenta lines). The cyclotron frequency was calculated from equation (\ref{eqn:freqbrelation1}) assuming a dipolar magnetic field, using the predicted magnetic moments of 0.76 $\mathcal M_{\rm J}$ from \citet{Griessmeier2007c} (dashed green line, denoted by G2007), the predicted 58 G (7.5 $\mathcal M_{\rm J}$) from \citet{Reiners2010} (dashed blue line, denoted by RC2010) and the value of 2.8 $\mathcal M_{\rm J}$ found in \citet{Griessmeier2011} (dashed magenta line, denoted by G2011). Radio emission is generated at frequencies close to the local cyclotron frequency of the electrons. Only if the cyclotron frequency exceeds the local plasma frequency ($f_{\rm p}/f_{\rm c} \leq 0.4$; \citep{Griessmeier2007c}), the condition for the generation of radio waves is fulfilled and radio waves with a frequency $f_{\rm c} > f_{\rm p}$ can be generated. Radio waves generated at a hypothetical distance $R_1$ can escape from their generation region through the planetary atmosphere/ionosphere if and only if the cyclotron frequency at the distance $R_1$ is higher than the plasma frequency for all distances $R_1 < R < \infty$.

All curves in Figure \ref{fig:fig3tau1} stop at the exobase level. The range of the plot stops at the exobase level for 0.046 AU, i.e.~$1.18 R_{\rm{p}}$. In the regions up to the exobase level for the magnetic field cases RC2010 and G2011 possibly generated radio waves can escape but not for G2007. However, the exobase levels are very low, i.e.~$1.18 R_{\rm{p}}$, $1.14 R_{\rm{p}}$ and $1.11 R_{\rm{p}}$ for 0.046, 0.1 and 0.2 AU, respectively. This means that the atmosphere is very compact. Beyond the exobase the conditions for the CMI are very likely much better. Thus, each magnetic field case should yield the possibility of generation and escape for radio emission. Only for the case G2007 emission generated in the small region below the exobase levels would not be able to escape. Beyond the exobase, all cyclotron frequencies should exceed the plasma frequencies. 

\begin{figure*}
	\includegraphics[width=0.8\columnwidth]{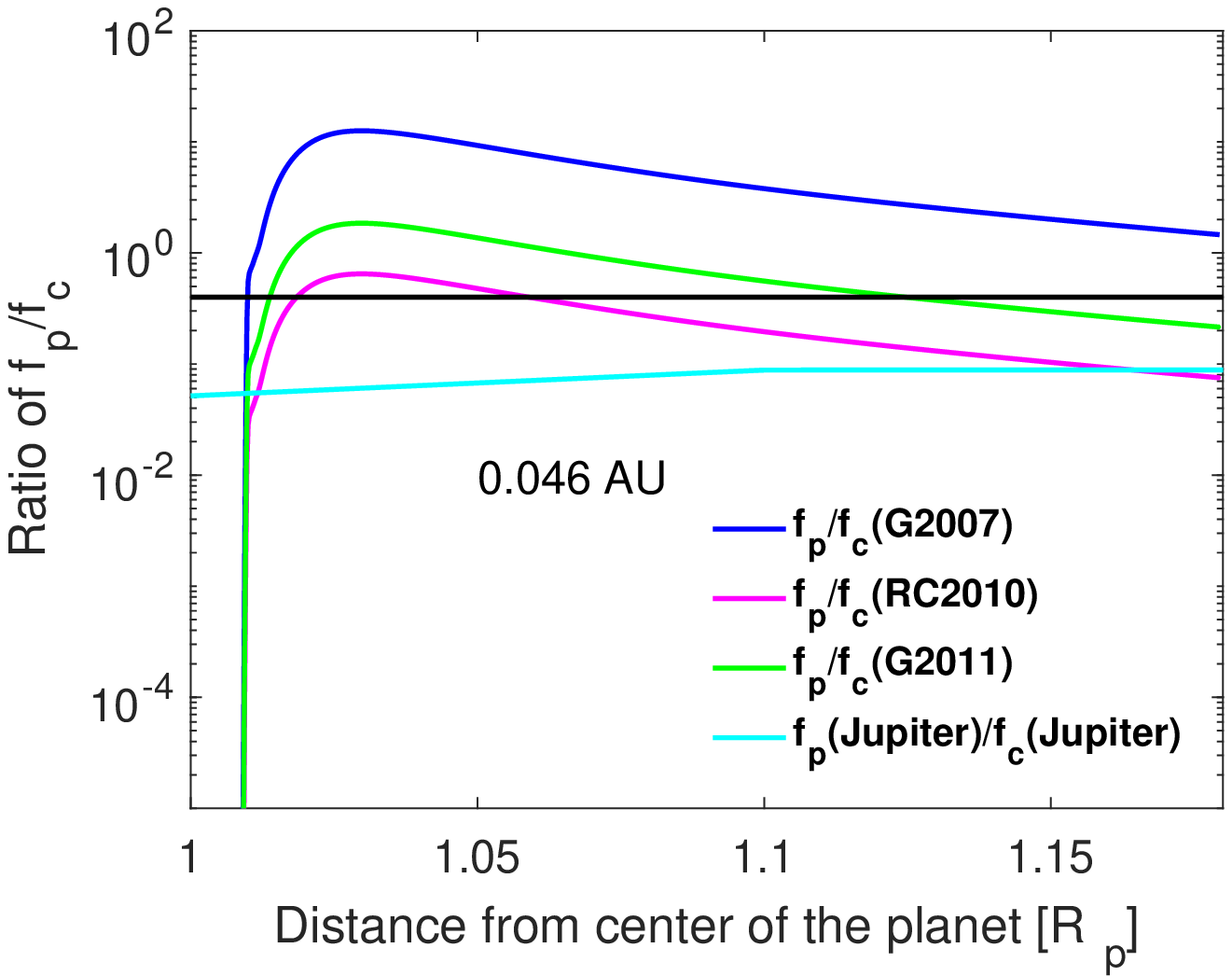}
\includegraphics[width=0.8\columnwidth]{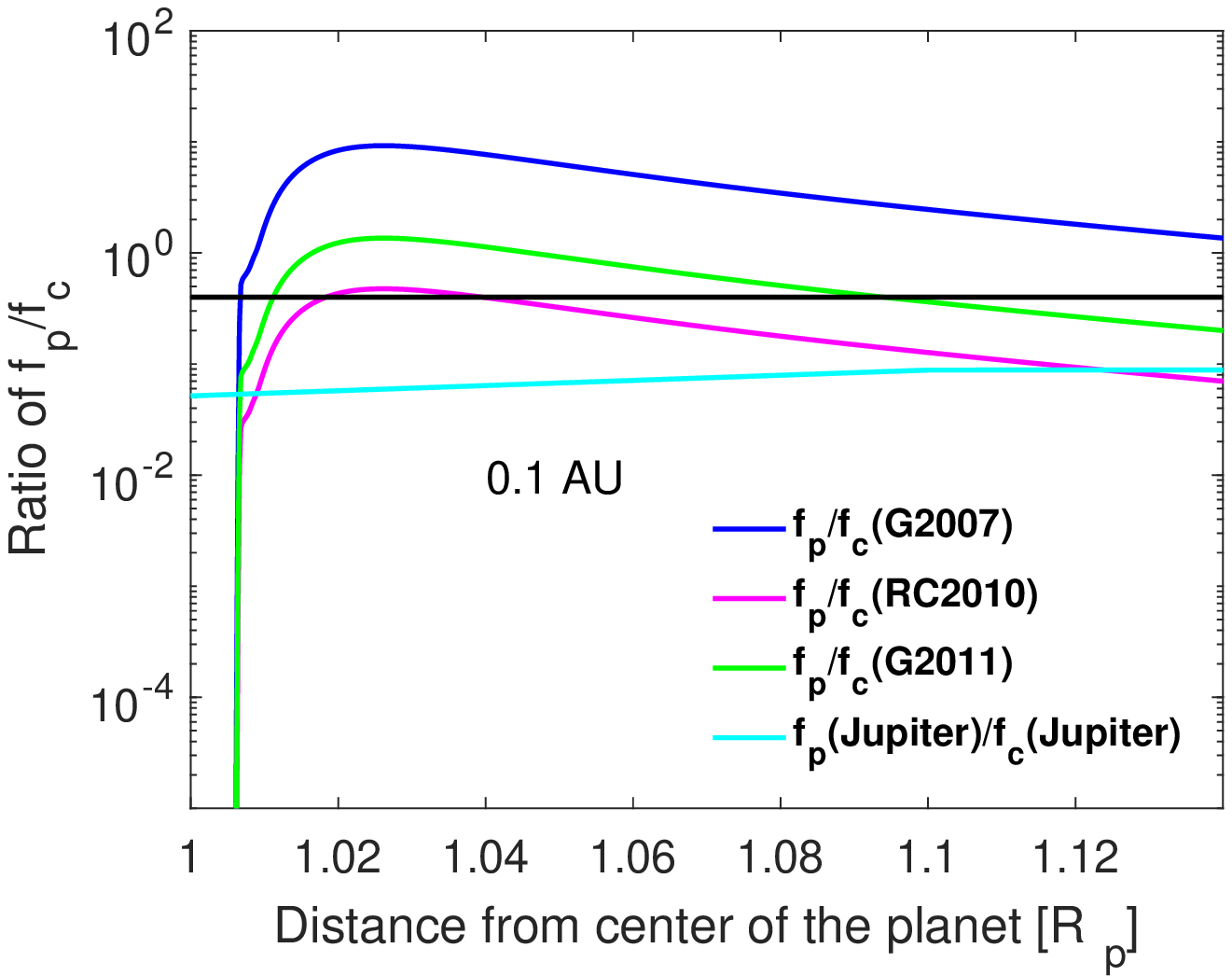}
\includegraphics[width=0.8\columnwidth]{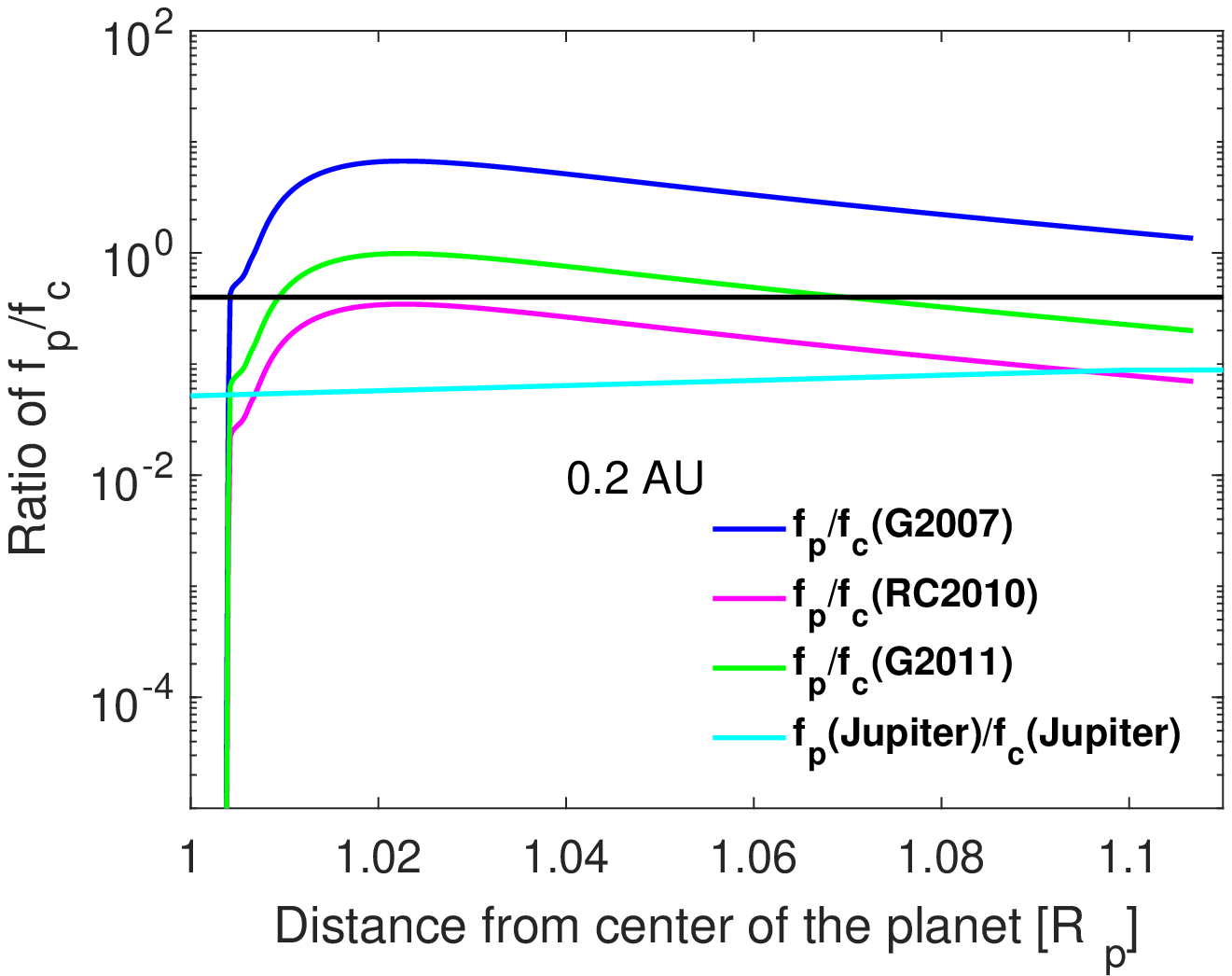}
    \caption{Ratio of plasma to cyclotron frequency for Tau Bootis b for different orbital distances. Upper-left panel: 0.046 AU. Upper-right panel: 0.1 AU. Lower panel: 0.2 AU. The black solid line indicates the maximum ratio of 0.4 for the CMI to work. Note that the curves end at the exobase levels.}
    \label{fig:fig4tau1}
\end{figure*}

Figure \ref{fig:fig4tau1} shows the ratios of plasma to cyclotron frequency for the orbits 0.046, 0.1 and 0.2 AU, respectively. For each orbit the corresponding plasma frequency is compared with the different cyclotron frequencies corresponding to the magnetic field cases as discussed above, i.e. the case G2007, corresponding to the magnetic moment as predicted by \citet{Griessmeier2007c}, the case RC2010, corresponding to the magnetic moment as predicted by \citet{Reiners2010} and the case G2011, corresponding to the magnetic moment as predicted by \citet{Griessmeier2011}. The solid black line indicates the maximum frequency ratio of 0.4 for the CMI to work \citep[see][]{Griessmeier2007b, Weber2017}. The same argument as above holds, i.e. because the plotting range stops at the exobase levels, the conditions for the CMI in these small region are not good only for the case G2007. Beyond the exobase, for every magnetic field case, the plasma frequency should be below the cyclotron frequency almost everywhere within the magnetosphere. This can be seen very well by the decreasing trend of the frequency ratios for all cases. When there is no hydrodynamic atmospheric outflow then exobase densities are always similar. Thus, we can expect a situation comparable to Jupiter in the solar system at 5.2 AU, where the frequency ratios are well below the indicated maximum of 0.4 because the plasma density decreases very fast beyond the exobase level. For details on the Jovian plasma environment we refer to e.g.~\citet{Prasad1971,Machida1978,Luhmann1980,Yelle1996,Stallard2001}.   

\begin{table*}
	\centering
	\caption{Summary of possibility for generation and escape of radio emission for the case studied in this paper. The case of Jupiter with $1 \mathcal{M}_{\rm J}$ is shown for comparison. The numbers in brackets are the maximum emission frequencies in MHz at the pole.}
	\label{tab:sum1}
	\begin{tabular}{p{3 cm}|p{3 cm}|p{3 cm}|p{3 cm}} 
		Magnetic moment [$\mathcal{M}_{\rm J}$] & 0.046 AU & 0.1 AU & 0.2 AU \\
		\hline
		\hline
		 & Tau Bootis & & \\
		\hline
		\hline
		$0.76 \mathcal{M}_{\rm J}$ ($\hat{=} 16.3$ MHz, G2007) & generation: $+$ & generation: $+$ & generation: $+$ \\
		 & escape: $+$ & escape: $+$ & escape: $+$ \\
		\hline
		$1 \mathcal{M}_{\rm J}$ ($\hat{=} 21.5$ MHz) & generation: $+$ & generation: $+$ & generation: $+$ \\
		 & escape: $+$ & escape: $+$ & escape: $+$ \\
		\hline
		$2.8 \mathcal{M}_{\rm J}$ ($\hat{=} 60$ MHz, G2011) &  generation: $+$ & generation: $+$ & generation: $+$ \\
		 & escape: $+$ & escape: $+$ & escape: $+$ \\
		\hline
		$7.5 \mathcal{M}_{\rm J}$ ($\hat{=} 161$ MHz, RC2010) &  generation: $+$ & generation: $+$ & generation: $+$ \\
		 & escape: $+$ & escape: $+$ & escape: $+$ \\
		\hline
		\hline
	\end{tabular}
\end{table*}

Table \ref{tab:sum1} gives a summary of the possibility for generation and escape of potential radio emission for the cases studied in this paper. The $+$ signs indicate that generation or escape is possible. The case of Jupiter ($1 \mathcal{M}_{\rm J}$) is intended as comparison to the cases studied here. To compare the results of the current study with our former findings, Table \ref{tab:tabsummm} shows the same summary as for Tau Bootis b, but for the planets HD 189733b and HD 209458b from \citet{Weber2017}. In the version of the table used here we filled the gaps in the table of \citet{Weber2017}.

\begin{table*}
\centering
\caption{Summary of possibility for generation and/or escape of radio waves for the planets HD 209458b and HD 189733b \citep{Weber2017}. The numbers in brackets are the maximum emission frequencies in MHz at the pole, assuming a planetary radius of $1.06 R_{\rm{J}}$.}
\label{tab:tabsummm}
\small{
\begin{tabular}{|p{4 cm}|p{2 cm}|p{2 cm}|p{2 cm}|p{2 cm}|}
\hline
 & HD 209458b (pole) & HD 209458b (equator) & HD 189733b (equator) & HD 209458b (1 AU) \\
\hline
\hline
$0.06 \mathcal{M}_{\rm J}$ ($\hat{=} 1.3$ MHz) & generation: $+$ (very close) & generation: $-$ & generation: $-$ & generation: $+$ (very close?) \\
 & escape: $-$ & escape: $-$ & escape: $-$ & escape: $-$? \\
\hline
$0.1 \mathcal{M}_{\rm J}$ ($\hat{=} 2.1$ MHz) & generation: $+$ (very close) & generation: $-$ & generation: $-$ & generation: $+$ (very close?) \\
 & escape: $-$ & escape: $-$ & escape: $-$ & escape: $-$? \\
\hline
$0.2 \mathcal{M}_{\rm J}$ ($\hat{=} 4.3$ MHz) & generation: $+$ (very close) & generation: $-$ & generation: $-$ & generation: $+$ (very close?) \\
 & escape: $-$ & escape: $-$ & escape: $-$ & escape: $-$? \\
\hline
$0.3 \mathcal{M}_{\rm J}$ ($\hat{=} 6.4$ MHz) & generation: $+$ (very close) & generation: $+$ (very close) & generation: $-$ & generation: $+$ (very close?)\\
 & escape: $-$ & escape: $-$ & escape: $-$ & escape: $-$? \\
\hline
$0.6 \mathcal{M}_{\rm J}$ ($\hat{=} 12.9$ MHz) & generation: $+$ (very close) & generation: $-$ & generation: $+$ (very close) & generation: $+$ (very close?) \\
 & escape: $-$ & escape: $-$ & escape: $-$ & escape: $-$? \\
\hline
$\mathcal{M}_{\rm J}$ ($\hat{=}21.5$ MHz) & generation: $+$ (very close) & generation: $+$ (very close) & generation: $+$ (very close) & generation: $+$ (very close?)\\
 & escape: $-$ & escape: $-$ & escape: $-$ & escape: $-$?\\
\hline
$5 \mathcal{M}_{\rm J}$ ($\hat{=}107.3$ MHz) & generation: $+$ & generation: $+$ & generation: $+$ & generation: $+$\\
 & escape: $-$ & escape: $-$ & escape: $-$ & escape: $+$ \\
\hline
$50 \mathcal{M}_{\rm J}$ ($\hat{=} 1073$ MHz) & generation: $+$ & generation: $+$ & generation: $+$ & generation: $+$\\
 & escape: $+$ & escape: $+$ & escape: $-$ & escape: $+$ \\
\hline
$100 \mathcal{M}_{\rm J}$ ($\hat{=} 2146$ MHz) & generation: $+$ & generation: $+$ & generation: $+$ & generation: $+$\\
 & escape: $+$ & escape: $+$ & escape: $+$ & escape: $+$ \\
\hline
\hline
\end{tabular}}
\end{table*}



%
%


%

\section{Implications for Future Observation Campaigns Targeting Tau Bootis b}\label{sec:sec5}
 
Radio emission from the Tau Bootis b system hasn't been detected yet, even though we find, in agreement with several former studies \citep[e.g.][]{Farrell2003, Lazio2007, Griessmeier2005, Griessmeier2006, Griessmeier2007a, Griessmeier2007b, Griessmeier2007c, Griessmeier2011}, that it should be a good candidate for successful observations. Radio fluxes were predicted to be in the sensitivity range of several radio telescopes, e.g.~UTR-2 and LOFAR. \citet{Griessmeier2011} compared the rotation independent model of \citet{Reiners2010} to their model where the planetary magnetic moment is dependent on rotation, and thus tidal locking has an influence. They find radio fluxes of 180 and 300 mJy for the rotation independent and the rotation dependent model, respectively. Since the cyclotron frequencies calculated in this paper are based on the magnetic moment predictions from these models, the maximum emission frequency and the radio flux densities are the same. Thus, we conclude that radio emission from Tau Bootis b should be detectable with LOFAR, UTR-2, the upcoming SKA (Square Kilometer Array), NenuFAR (New Extension in Nan\c{c}ay Upgrading LOFAR, currently under construction) \citep{Griessmeier2011} and GURT \citep[under construction in Kharkov, Ukraine][]{Konovalenko2016}. Also the VLA (Very Large Array), the LWA (Long Wavelength Array) and the GMRT (Giant Metrewave Radio Telescope) should be sensitive enough to allow a detection \citep{Griessmeier2011}.

Finally, we note that for HD 209458b radiative cooling can be neglected, whereas for HD 189733b it already has an effect. For details we refer to the study by \citet{Salz2015}, who described HD 189733b as an intermediate case. We plan to perform a similar study as for Tau Bootis b for HD 189733b. Recently, \citet{Lalitha2018} studied the atmospheric mass loss of four close-in planets around the very active stars Kepler-63, Kepler-210, WASP-19, and HAT-P-11. They found the XUV luminosities of these stars to be orders of magnitude higher than for the Sun. \citet{Lalitha2018} compared the four studied planets with HD 209458b and HD 189733b. All planets suffer extreme atmospheric mass loss due to the strong XUV radiation from their host stars, very likely leading to the extended upper atmospheres and ionospheres, which might inhibit the escape of possibly generated radio emission.

\section{Conclusions}\label{sec:sec6}

We find that the supermassive Hot Jupiter Tau Bootis b is clearly more favourable for the CMI than hot Jupiters like HD 209458b or HD 189733b studied previously. The main issue with the latter planets is the extended upper atmosphere and ionosphere, where exobase distances exceed the magnetospheric standoff distance, and upper atmospheres are in a hydrodynamic state for orbits $<$ 0.2 AU for Sun-like stars and $<$ 0.5 AU for more active stars \citep{Weber2017}.

In general, we find that supermassive hot Jupiters like Tau Bootis b are much better candidates for future radio observations than classical hot Jupiters. There is no hydrodynamic outflow and the conditions for the CMI are very good, especially if compared to less massive hot Jupiters like HD 209458b or HD 189733b (0.69 and 1.142 Jupiter masses, respectively, from \url{http://exoplanet.eu}, accessed 01.06.2018), where the atmospheric outflow is hydrodynamic. It is also  worth to note that recently \citet{Yates2017,Yates2018} found the same result for "normal" hot Jupiters as in \citet{Weber2017,Weber2017b}.

Considering our example planet Tau Bootis b, we can definitely say that objects with 5.84 $M_{\rm J}$ (Jupiter masses) keep their atmosphere compact due to their large gravity and thus lead to favourable conditions for the CMI. In follow-up studies we will investigate at which mass the transition to these conditions starts. This will include variations of the type of the star or the planetary radius. A further step in our follow-up studies will include an extension of the investigated magnetospheric regions beyond the exobase level up to the magnetopause. 

\section*{Acknowledgements}

C. Weber, H. Lammer and P. Odert acknowledge support from the FWF project P27256-N27 `Characterizing Stellar and Exoplanetary Environments via Modeling of Lyman-$\alpha$ Transit Observations of hot Jupiters'. P. Odert also acknowledges the Austrian Science Fund (FWF): P30949-N36. The authors acknowledge also the support by the FWF NFN projects S11606-N16 `Magnetospheres - Magnetospheric Electrodynamics of Exoplanets' and S11607-N16 `Particle/Radiative Interactions with Upper Atmospheres of Planetary Bodies Under Extreme Stellar Conditions'. N. Erkaev and V. Ivanov acknowledge support by the Russian Science Foundation grant No 18-12-00080. The authors also thank an anonymous referee for valuable comments and suggestions which helped to improve this paper.





\bibliographystyle{mnras}
\bibliography{mybibtex} 

\begin{thebibliography}{}
\makeatletter
\relax
\def\mn@urlcharsother{\let\do\@makeother \do\$\do\&\do\#\do\^\do\_\do\%\do\~}
\def\mn@doi{\begingroup\mn@urlcharsother \@ifnextchar [ {\mn@doi@}
  {\mn@doi@[]}}
\def\mn@doi@[#1]#2{\def\@tempa{#1}\ifx\@tempa\@empty \href
  {http://dx.doi.org/#2} {doi:#2}\else \href {http://dx.doi.org/#2} {#1}\fi
  \endgroup}
\def\mn@eprint#1#2{\mn@eprint@#1:#2::\@nil}
\def\mn@eprint@arXiv#1{\href {http://arxiv.org/abs/#1} {{\tt arXiv:#1}}}
\def\mn@eprint@dblp#1{\href {http://dblp.uni-trier.de/rec/bibtex/#1.xml}
  {dblp:#1}}
\def\mn@eprint@#1:#2:#3:#4\@nil{\def\@tempa {#1}\def\@tempb {#2}\def\@tempc
  {#3}\ifx \@tempc \@empty \let \@tempc \@tempb \let \@tempb \@tempa \fi \ifx
  \@tempb \@empty \def\@tempb {arXiv}\fi \@ifundefined
  {mn@eprint@\@tempb}{\@tempb:\@tempc}{\expandafter \expandafter \csname
  mn@eprint@\@tempb\endcsname \expandafter{\@tempc}}}

\bibitem[\protect\citeauthoryear{{Bastian}, {Dulk}  \& {Leblanc}}{{Bastian}
  et~al.}{2000}]{Bastian2000}
{Bastian} T.~S.,  {Dulk} G.~A.,   {Leblanc} Y.,  2000, \mn@doi [\apj]
  {10.1086/317864}, \href {http://adsabs.harvard.edu/abs/2000ApJ...545.1058B}
  {545, 1058}

\bibitem[\protect\citeauthoryear{{Berger} et~al.,}{{Berger}
  et~al.}{2010}]{Berger2010}
{Berger} E.,  et~al., 2010, \mn@doi [\apj] {10.1088/0004-637X/709/1/332}, \href
  {http://adsabs.harvard.edu/abs/2010ApJ...709..332B} {709, 332}

\bibitem[\protect\citeauthoryear{{Black}}{{Black}}{1981}]{Black1981}
{Black} J.~H.,  1981, \mn@doi [\mnras] {10.1093/mnras/197.3.553}, \href
  {http://adsabs.harvard.edu/abs/1981MNRAS.197..553B} {197, 553}

\bibitem[\protect\citeauthoryear{{Cubillos}}{{Cubillos}}{2016}]{Cubillos2015}
{Cubillos} P.~E.,  2016, PhD thesis, University of Central Florida (\mn@eprint
  {arXiv} {1604.01320})

\bibitem[\protect\citeauthoryear{{Cubillos} et~al.,}{{Cubillos}
  et~al.}{2017}]{Cubillos2017}
{Cubillos} P.,  et~al., 2017, \mn@doi [\mnras] {10.1093/mnras/stw3103}, \href
  {http://adsabs.harvard.edu/abs/2017MNRAS.466.1868C} {466, 1868}

\bibitem[\protect\citeauthoryear{{Daley-Yates} \& {Stevens}}{{Daley-Yates} \&
  {Stevens}}{2017}]{Yates2017}
{Daley-Yates} S.,  {Stevens} I.~R.,  2017, \mn@doi [Astronomische Nachrichten]
  {10.1002/asna.201713395}, \href
  {http://adsabs.harvard.edu/abs/2017AN....338..881D} {338, 881}

\bibitem[\protect\citeauthoryear{{Daley-Yates} \& {Stevens}}{{Daley-Yates} \&
  {Stevens}}{2018}]{Yates2018}
{Daley-Yates} S.,  {Stevens} I.~R.,  2018, preprint, \href
  {http://adsabs.harvard.edu/abs/2018arXiv180608147D} {} (\mn@eprint {arXiv}
  {1806.08147})

\bibitem[\protect\citeauthoryear{{Donati} et~al.,}{{Donati}
  et~al.}{2008}]{Donati2008}
{Donati} J.-F.,  et~al., 2008, \mn@doi [\mnras]
  {10.1111/j.1365-2966.2008.12946.x}, \href
  {http://adsabs.harvard.edu/abs/2008MNRAS.385.1179D} {385, 1179}

\bibitem[\protect\citeauthoryear{{Ergun}, {Carlson}, {McFadden}, {Delory},
  {Strangeway}  \& {Pritchett}}{{Ergun} et~al.}{2000}]{Ergun2000}
{Ergun} R.~E.,  {Carlson} C.~W.,  {McFadden} J.~P.,  {Delory} G.~T.,
  {Strangeway} R.~J.,   {Pritchett} P.~L.,  2000, \mn@doi [\apj]
  {10.1086/309094}, \href {http://adsabs.harvard.edu/abs/2000ApJ...538..456E}
  {538, 456}

\bibitem[\protect\citeauthoryear{{Erkaev}, {Lammer}, {Odert}, {Kulikov}  \&
  {Kislyakova}}{{Erkaev} et~al.}{2015}]{Erkaev2015}
{Erkaev} N.~V.,  {Lammer} H.,  {Odert} P.,  {Kulikov} Y.~N.,   {Kislyakova}
  K.~G.,  2015, \mn@doi [\mnras] {10.1093/mnras/stv130}, \href
  {http://adsabs.harvard.edu/abs/2015MNRAS.448.1916E} {448, 1916}

\bibitem[\protect\citeauthoryear{{Erkaev}, {Lammer}, {Odert}, {Kislyakova},
  {Johnstone}, {G{\"u}del}  \& {Khodachenko}}{{Erkaev}
  et~al.}{2016}]{Erkaev2016}
{Erkaev} N.~V.,  {Lammer} H.,  {Odert} P.,  {Kislyakova} K.~G.,  {Johnstone}
  C.~P.,  {G{\"u}del} M.,   {Khodachenko} M.~L.,  2016, \mn@doi [\mnras]
  {10.1093/mnras/stw935}, \href {http://esoads.eso.org/abs/2016MNRAS.460.1300E}
  {460, 1300}

\bibitem[\protect\citeauthoryear{{Fares} et~al.,}{{Fares}
  et~al.}{2010}]{Fares2010}
{Fares} R.,  et~al., 2010, \mn@doi [\mnras] {10.1111/j.1365-2966.2010.16715.x},
  \href {http://adsabs.harvard.edu/abs/2010MNRAS.406..409F} {406, 409}

\bibitem[\protect\citeauthoryear{{Farrell}, {Desch}  \& {Zarka}}{{Farrell}
  et~al.}{1999}]{Farrell1999}
{Farrell} W.~M.,  {Desch} M.~D.,   {Zarka} P.,  1999, \mn@doi [\jgr]
  {10.1029/1998JE900050}, \href
  {http://adsabs.harvard.edu/abs/1999JGR...10414025F} {104, 14025}

\bibitem[\protect\citeauthoryear{{Farrell}, {Desch}, {Lazio}, {Bastian}  \&
  {Zarka}}{{Farrell} et~al.}{2003}]{Farrell2003}
{Farrell} W.~M.,  {Desch} M.~D.,  {Lazio} T.~J.,  {Bastian} T.,   {Zarka} P.,
  2003, in {Deming} D.,  {Seager} S.,  eds,  Astronomical Society of the
  Pacific Conference Series Vol. 294, Scientific Frontiers in Research on
  Extrasolar Planets. pp 151--156

\bibitem[\protect\citeauthoryear{{Fossati}, {France}, {Koskinen}, {Juvan},
  {Haswell}  \& {Lendl}}{{Fossati} et~al.}{2015}]{Fossati2015}
{Fossati} L.,  {France} K.,  {Koskinen} T.,  {Juvan} I.~G.,  {Haswell} C.~A.,
  {Lendl} M.,  2015, \mn@doi [\apj] {10.1088/0004-637X/815/2/118}, \href
  {http://adsabs.harvard.edu/abs/2015ApJ...815..118F} {815, 118}

\bibitem[\protect\citeauthoryear{{Fossati} et~al.,}{{Fossati}
  et~al.}{2017}]{Fossati2017}
{Fossati} L.,  et~al., 2017, \mn@doi [\aap] {10.1051/0004-6361/201629716},
  \href {http://adsabs.harvard.edu/abs/2017A%26A...598A..90F} {598, A90}

\bibitem[\protect\citeauthoryear{{Fossati}, {Koskinen}, {France}, {Cubillos},
  {Haswell}, {Lanza}  \& {Pillitteri}}{{Fossati} et~al.}{2018}]{Fossati2018}
{Fossati} L.,  {Koskinen} T.,  {France} K.,  {Cubillos} P.~E.,  {Haswell}
  C.~A.,  {Lanza} A.~F.,   {Pillitteri} I.,  2018, \mn@doi [\aj]
  {10.3847/1538-3881/aaa891}, \href
  {http://adsabs.harvard.edu/abs/2018AJ....155..113F} {155, 113}

\bibitem[\protect\citeauthoryear{{Glover} \& {Jappsen}}{{Glover} \&
  {Jappsen}}{2007}]{Glover2007}
{Glover} S.~C.~O.,  {Jappsen} A.-K.,  2007, \mn@doi [\apj] {10.1086/519445},
  \href {http://adsabs.harvard.edu/abs/2007ApJ...666....1G} {666, 1}

\bibitem[\protect\citeauthoryear{{Grie{\ss}meier}}{{Grie{\ss}meier}}{2007}]{Griessmeier2007a}
{Grie{\ss}meier} J.-M.,  2007, \mn@doi [\planss] {10.1016/j.pss.2006.11.003},
  \href {http://adsabs.harvard.edu/abs/2007P%26SS...55..530G} {55, 530}

\bibitem[\protect\citeauthoryear{{Grie{\ss}meier}}{{Grie{\ss}meier}}{2017}]{Griessmeier2017}
{Grie{\ss}meier} J.-M.,  2017, in {Fischer} G. e.~a.,  ed., Planetary Radio
  Emission VIII. Austrian Academy of Sciences Press, Vienna, in press.
pp 285--299

\bibitem[\protect\citeauthoryear{{Grie{\ss}meier} et~al.,}{{Grie{\ss}meier}
  et~al.}{2004}]{Griessmeier2004}
{Grie{\ss}meier} J.-M.,  et~al., 2004, \mn@doi [\aap]
  {10.1051/0004-6361:20035684}, \href
  {http://adsabs.harvard.edu/abs/2004A%26A...425..753G} {425, 753}

\bibitem[\protect\citeauthoryear{{Grie{\ss}meier}, {Motschmann}, {Mann}  \&
  {Rucker}}{{Grie{\ss}meier} et~al.}{2005}]{Griessmeier2005}
{Grie{\ss}meier} J.-M.,  {Motschmann} U.,  {Mann} G.,   {Rucker} H.~O.,  2005,
  \mn@doi [\aap] {10.1051/0004-6361:20041976}, \href
  {http://adsabs.harvard.edu/abs/2005A%26A...437..717G} {437, 717}

\bibitem[\protect\citeauthoryear{Grie{\ss}meier, Motschmann, Glassmeier, Mann
  \& Rucker}{Grie{\ss}meier et~al.}{2006a}]{Griessmeier2006b}
Grie{\ss}meier J.,  Motschmann U.,  Glassmeier K.,  Mann G.,   Rucker H.,
  2006a, in Tenth Anniversary of 51 Peg-b: Status of and prospects for hot
  Jupiter studies. pp 259--266

\bibitem[\protect\citeauthoryear{{Grie{\ss}meier}, {Motschmann}, {Khodachenko}
  \& {Rucker}}{{Grie{\ss}meier} et~al.}{2006b}]{Griessmeier2006}
{Grie{\ss}meier} J.-M.,  {Motschmann} U.,  {Khodachenko} M.,   {Rucker} H.~O.,
  2006b, in {Rucker} H.~O.,  {Kurth} W.,   {Mann} G.,  eds, Planetary Radio
  Emissions VI. p.~571

\bibitem[\protect\citeauthoryear{{Grie{\ss}meier}, {Preusse}, {Khodachenko},
  {Motschmann}, {Mann}  \& {Rucker}}{{Grie{\ss}meier}
  et~al.}{2007a}]{Griessmeier2007c}
{Grie{\ss}meier} J.-M.,  {Preusse} S.,  {Khodachenko} M.,  {Motschmann} U.,
  {Mann} G.,   {Rucker} H.~O.,  2007a, \mn@doi [\planss]
  {10.1016/j.pss.2006.01.008}, \href
  {http://adsabs.harvard.edu/abs/2007P%26SS...55..618G} {55, 618}

\bibitem[\protect\citeauthoryear{{Grie{\ss}meier}, {Zarka}  \&
  {Spreeuw}}{{Grie{\ss}meier} et~al.}{2007b}]{Griessmeier2007b}
{Grie{\ss}meier} J.-M.,  {Zarka} P.,   {Spreeuw} H.,  2007b, \mn@doi [\aap]
  {10.1051/0004-6361:20077397}, \href
  {http://adsabs.harvard.edu/abs/2007A%26A...475..359G} {475, 359}

\bibitem[\protect\citeauthoryear{{Grie{\ss}meier}, {Zarka}  \&
  {Girard}}{{Grie{\ss}meier} et~al.}{2011}]{Griessmeier2011}
{Grie{\ss}meier} J.-M.,  {Zarka} P.,   {Girard} J.~N.,  2011, \mn@doi [Radio
  Science] {10.1029/2011RS004752}, \href
  {http://adsabs.harvard.edu/abs/2011RaSc...46.0F09G} {46, 0}

\bibitem[\protect\citeauthoryear{Hallinan, Antonova, Doyle, Bourke, Lane  \&
  Golden}{Hallinan et~al.}{2008}]{Hallinan2008}
Hallinan G.,  Antonova A.,  Doyle J.,  Bourke S.,  Lane C.,   Golden A.,  2008,
  The Astrophysical Journal, 684, 644

\bibitem[\protect\citeauthoryear{{Hallinan}, {Sirothia}, {Antonova},
  {Ishwara-Chandra}, {Bourke}, {Doyle}, {Hartman}  \& {Golden}}{{Hallinan}
  et~al.}{2013}]{Hallinan2012}
{Hallinan} G.,  {Sirothia} S.~K.,  {Antonova} A.,  {Ishwara-Chandra} C.~H.,
  {Bourke} S.,  {Doyle} J.~G.,  {Hartman} J.,   {Golden} A.,  2013, \mn@doi
  [\apj] {10.1088/0004-637X/762/1/34}, \href
  {http://adsabs.harvard.edu/abs/2013ApJ...762...34H} {762, 34}

\bibitem[\protect\citeauthoryear{{Hallinan} et~al.,}{{Hallinan}
  et~al.}{2015}]{Hallinan2015}
{Hallinan} G.,  et~al., 2015, \mn@doi [\nat] {10.1038/nature14619}, \href
  {http://adsabs.harvard.edu/abs/2015Natur.523..568H} {523, 568}

\bibitem[\protect\citeauthoryear{{Herrero}, {Morales}, {Ribas}  \&
  {Naves}}{{Herrero} et~al.}{2011}]{Herrero2011}
{Herrero} E.,  {Morales} J.~C.,  {Ribas} I.,   {Naves} R.,  2011, \mn@doi
  [\aap] {10.1051/0004-6361/201015875}, \href
  {http://adsabs.harvard.edu/abs/2011A%26A...526L..10H} {526, L10}

\bibitem[\protect\citeauthoryear{{Khodachenko} et~al.,}{{Khodachenko}
  et~al.}{2012}]{Khodachenko2012}
{Khodachenko} M.~L.,  et~al., 2012, \mn@doi [\apj]
  {10.1088/0004-637X/744/1/70}, \href
  {http://adsabs.harvard.edu/abs/2012ApJ...744...70K} {744, 70}

\bibitem[\protect\citeauthoryear{Kislyakova, Holmstr{\"o}m, Lammer, Odert  \&
  Khodachenko}{Kislyakova et~al.}{2014}]{Kislyakova2014}
Kislyakova K.~G.,  Holmstr{\"o}m M.,  Lammer H.,  Odert P.,   Khodachenko
  M.~L.,  2014, Science, 346, 981

\bibitem[\protect\citeauthoryear{{Konovalenko} et~al.,}{{Konovalenko}
  et~al.}{2016}]{Konovalenko2016}
{Konovalenko} A.,  et~al., 2016, \mn@doi [Experimental Astronomy]
  {10.1007/s10686-016-9498-x}, \href
  {http://adsabs.harvard.edu/abs/2016ExA....42...11K} {42, 11}

\bibitem[\protect\citeauthoryear{{Lalitha}, {Schmitt}  \& {Dash}}{{Lalitha}
  et~al.}{2018}]{Lalitha2018}
{Lalitha} S.,  {Schmitt} J.~H.~M.~M.,   {Dash} S.,  2018, \mn@doi [\mnras]
  {10.1093/mnras/sty732}, \href
  {http://adsabs.harvard.edu/abs/2018MNRAS.477..808L} {477, 808}

\bibitem[\protect\citeauthoryear{{Lammer} et~al.,}{{Lammer}
  et~al.}{2016}]{Lammer2016}
{Lammer} H.,  et~al., 2016, \mn@doi [\mnras] {10.1093/mnrasl/slw095}, \href
  {http://adsabs.harvard.edu/abs/2016MNRAS.461L..62L} {461, L62}

\bibitem[\protect\citeauthoryear{{Lazio} \& {Farrell}}{{Lazio} \&
  {Farrell}}{2007}]{Lazio2007}
{Lazio} T.~J.~W.,  {Farrell} W.~M.,  2007, \mn@doi [\apj] {10.1086/519730},
  \href {http://adsabs.harvard.edu/abs/2007ApJ...668.1182L} {668, 1182}

\bibitem[\protect\citeauthoryear{{Lazio}, {Farrell}, {Dietrick}, {Greenlees},
  {Hogan}, {Jones}  \& {Hennig}}{{Lazio} et~al.}{2004}]{Lazio2004}
{Lazio} W. T.~J.,  {Farrell} W.~M.,  {Dietrick} J.,  {Greenlees} E.,  {Hogan}
  E.,  {Jones} C.,   {Hennig} L.~A.,  2004, \mn@doi [\apj] {10.1086/422449},
  \href {http://adsabs.harvard.edu/abs/2004ApJ...612..511L} {612, 511}

\bibitem[\protect\citeauthoryear{{Linsky}, {France}  \& {Ayres}}{{Linsky}
  et~al.}{2013}]{Linsky2013}
{Linsky} J.~L.,  {France} K.,   {Ayres} T.,  2013, \mn@doi [\apj]
  {10.1088/0004-637X/766/2/69}, \href
  {http://adsabs.harvard.edu/abs/2013ApJ...766...69L} {766, 69}

\bibitem[\protect\citeauthoryear{{Linsky}, {Fontenla}  \& {France}}{{Linsky}
  et~al.}{2014}]{Linsky2014}
{Linsky} J.~L.,  {Fontenla} J.,   {France} K.,  2014, \mn@doi [\apj]
  {10.1088/0004-637X/780/1/61}, \href
  {http://adsabs.harvard.edu/abs/2014ApJ...780...61L} {780, 61}

\bibitem[\protect\citeauthoryear{{Llama}, {Jardine}, {Wood}, {Hallinan}  \&
  {Morin}}{{Llama} et~al.}{2018}]{Llama2018}
{Llama} J.,  {Jardine} M.~M.,  {Wood} K.,  {Hallinan} G.,   {Morin} J.,  2018,
  \mn@doi [\apj] {10.3847/1538-4357/aaa59f}, \href
  {http://adsabs.harvard.edu/abs/2018ApJ...854....7L} {854, 7}

\bibitem[\protect\citeauthoryear{{Luhmann} \& {Walker}}{{Luhmann} \&
  {Walker}}{1980}]{Luhmann1980}
{Luhmann} J.~G.,  {Walker} R.~J.,  1980, \mn@doi [\icarus]
  {10.1016/0019-1035(80)90030-5}, \href
  {http://adsabs.harvard.edu/abs/1980Icar...44..361L} {44, 361}

\bibitem[\protect\citeauthoryear{{Lynch}, {Murphy}, {Lenc}  \&
  {Kaplan}}{{Lynch} et~al.}{2018}]{Lynch2018}
{Lynch} C.~R.,  {Murphy} T.,  {Lenc} E.,   {Kaplan} D.~L.,  2018, \mn@doi
  [\mnras] {10.1093/mnras/sty1138}, \href
  {http://adsabs.harvard.edu/abs/2018MNRAS.tmp.1077L} {}

\bibitem[\protect\citeauthoryear{{Machida} \& {Nishida}}{{Machida} \&
  {Nishida}}{1978}]{Machida1978}
{Machida} S.,  {Nishida} A.,  1978, \mn@doi [\planss]
  {10.1016/0032-0633(78)90005-3}, \href
  {http://adsabs.harvard.edu/abs/1978P%26SS...26..745M} {26, 745}

\bibitem[\protect\citeauthoryear{{Majid}, {Winterhalter}, {Chandra}, {Kuiper},
  {Lazio}, {Naudet}  \& {Zarka}}{{Majid} et~al.}{2006}]{Majid2006}
{Majid} W.,  {Winterhalter} D.,  {Chandra} I.,  {Kuiper} T.,  {Lazio} J.,
  {Naudet} C.,   {Zarka} P.,  2006, in {Rucker} H.~O.,  {Kurth} W.,   {Mann}
  G.,  eds, Planetary Radio Emissions VI. p.~589

\bibitem[\protect\citeauthoryear{{Murray-Clay}, {Chiang}  \&
  {Murray}}{{Murray-Clay} et~al.}{2009}]{MurrayClay2009}
{Murray-Clay} R.~A.,  {Chiang} E.~I.,   {Murray} N.,  2009, \mn@doi [\apj]
  {10.1088/0004-637X/693/1/23}, \href
  {http://adsabs.harvard.edu/abs/2009ApJ...693...23M} {693, 23}

\bibitem[\protect\citeauthoryear{{Prasad} \& {Capone}}{{Prasad} \&
  {Capone}}{1971}]{Prasad1971}
{Prasad} S.~S.,  {Capone} L.~A.,  1971, \mn@doi [\icarus]
  {10.1016/0019-1035(71)90032-7}, \href
  {http://adsabs.harvard.edu/abs/1971Icar...15...45P} {15, 45}

\bibitem[\protect\citeauthoryear{{Reiners} \& {Christensen}}{{Reiners} \&
  {Christensen}}{2010}]{Reiners2010}
{Reiners} A.,  {Christensen} U.~R.,  2010, \mn@doi [\aap]
  {10.1051/0004-6361/201014251}, \href
  {http://adsabs.harvard.edu/abs/2010A%26A...522A..13R} {522, A13}

\bibitem[\protect\citeauthoryear{{Route} \& {Wolszczan}}{{Route} \&
  {Wolszczan}}{2016a}]{Route2016a}
{Route} M.,  {Wolszczan} A.,  2016a, \mn@doi [\apjl]
  {10.3847/2041-8205/821/2/L21}, \href
  {http://adsabs.harvard.edu/abs/2016ApJ...821L..21R} {821, L21}

\bibitem[\protect\citeauthoryear{{Route} \& {Wolszczan}}{{Route} \&
  {Wolszczan}}{2016b}]{Route2016b}
{Route} M.,  {Wolszczan} A.,  2016b, \mn@doi [\apj]
  {10.3847/0004-637X/830/2/85}, \href
  {http://adsabs.harvard.edu/abs/2016ApJ...830...85R} {830, 85}

\bibitem[\protect\citeauthoryear{{Ryabov}, {Zarka}  \& {Ryabov}}{{Ryabov}
  et~al.}{2004}]{Ryabov2004}
{Ryabov} V.~B.,  {Zarka} P.,   {Ryabov} B.~P.,  2004, \mn@doi [\planss]
  {10.1016/j.pss.2004.09.019}, \href
  {http://adsabs.harvard.edu/abs/2004P%26SS...52.1479R} {52, 1479}

\bibitem[\protect\citeauthoryear{Salz, Czesla, Schneider  \& Schmitt}{Salz
  et~al.}{2015}]{Salz2015}
Salz M.,  Czesla S.,  Schneider P.,   Schmitt J.,  2015, arXiv preprint
  arXiv:1511.09341

\bibitem[\protect\citeauthoryear{{Shematovich}, {Ionov}  \&
  {Lammer}}{{Shematovich} et~al.}{2014}]{Shematovich2014}
{Shematovich} V.~I.,  {Ionov} D.~E.,   {Lammer} H.,  2014, \mn@doi [\aap]
  {10.1051/0004-6361/201423573}, \href
  {http://adsabs.harvard.edu/abs/2014A%26A...571A..94S} {571, A94}

\bibitem[\protect\citeauthoryear{{Shiratori}, {Yokoo}, {Saso}, {Kameya},
  {Iwadate}  \& {Asari}}{{Shiratori} et~al.}{2006}]{Shiratori2006}
{Shiratori} Y.,  {Yokoo} H.,  {Saso} T.,  {Kameya} O.,  {Iwadate} K.,   {Asari}
  K.,  2006, in {Arnold} L.,  {Bouchy} F.,   {Moutou} C.,  eds, Tenth
  Anniversary of 51 Peg-b: Status of and prospects for hot Jupiter studies. pp
  290--292

\bibitem[\protect\citeauthoryear{{Shulyak}, {Tsymbal}, {Ryabchikova},
  {St{\"u}tz}  \& {Weiss}}{{Shulyak} et~al.}{2004}]{Shulyak2004}
{Shulyak} D.,  {Tsymbal} V.,  {Ryabchikova} T.,  {St{\"u}tz} C.,   {Weiss}
  W.~W.,  2004, \mn@doi [\aap] {10.1051/0004-6361:20034169}, \href
  {http://adsabs.harvard.edu/abs/2004A%26A...428..993S} {428, 993}

\bibitem[\protect\citeauthoryear{{Stallard}, {Miller}, {Millward}  \&
  {Joseph}}{{Stallard} et~al.}{2001}]{Stallard2001}
{Stallard} T.,  {Miller} S.,  {Millward} G.,   {Joseph} R.~D.,  2001, \mn@doi
  [\icarus] {10.1006/icar.2001.6681}, \href
  {http://adsabs.harvard.edu/abs/2001Icar..154..475S} {154, 475}

\bibitem[\protect\citeauthoryear{{Storey} \& {Hummer}}{{Storey} \&
  {Hummer}}{1995}]{Storey1995}
{Storey} P.~J.,  {Hummer} D.~G.,  1995, \mn@doi [\mnras]
  {10.1093/mnras/272.1.41}, \href
  {http://adsabs.harvard.edu/abs/1995MNRAS.272...41S} {272, 41}

\bibitem[\protect\citeauthoryear{{Stroe}, {Snellen}  \&
  {R{\"o}ttgering}}{{Stroe} et~al.}{2012}]{Stroe2012}
{Stroe} A.,  {Snellen} I.~A.~G.,   {R{\"o}ttgering} H.~J.~A.,  2012, \mn@doi
  [\aap] {10.1051/0004-6361/201220006}, \href
  {http://adsabs.harvard.edu/abs/2012A%26A...546A.116S} {546, A116}

\bibitem[\protect\citeauthoryear{{Treumann}}{{Treumann}}{2006}]{Treumann2006}
{Treumann} R.~A.,  2006, \mn@doi [\aapr] {10.1007/s00159-006-0001-y}, \href
  {http://adsabs.harvard.edu/abs/2006A%26ARv..13..229T} {13, 229}

\bibitem[\protect\citeauthoryear{{Turner}, {Grie{\ss}meier}, {Zarka}  \&
  {Vasylieva}}{{Turner} et~al.}{2017}]{Turner2017b}
{Turner} J.,  {Grie{\ss}meier} J.-M.,  {Zarka} P.,   {Vasylieva} I.,  2017, in
  {Fischer} G. e.~a.,  ed., Planetary Radio Emission VIII. Austrian Academy of
  Sciences Press, Vienna, in press.
pp 301--313

\bibitem[\protect\citeauthoryear{{Watson}, {Donahue}  \& {Walker}}{{Watson}
  et~al.}{1981}]{Watson1981}
{Watson} A.~J.,  {Donahue} T.~M.,   {Walker} J.~C.~G.,  1981, \mn@doi [\icarus]
  {10.1016/0019-1035(81)90101-9}, \href
  {http://adsabs.harvard.edu/abs/1981Icar...48..150W} {48, 150}

\bibitem[\protect\citeauthoryear{{Weber} et~al.,}{{Weber}
  et~al.}{2017a}]{Weber2017b}
{Weber} C.,  et~al., 2017a, in {Fischer} G. e.~a.,  ed., Planetary Radio
  Emission VIII. Austrian Academy of Sciences Press, Vienna, in press.
pp 317--329

\bibitem[\protect\citeauthoryear{{Weber} et~al.,}{{Weber}
  et~al.}{2017b}]{Weber2017}
{Weber} C.,  et~al., 2017b, \mn@doi [\mnras] {10.1093/mnras/stx1099}, \href
  {http://adsabs.harvard.edu/abs/2017MNRAS.469.3505W} {469, 3505}

\bibitem[\protect\citeauthoryear{{Williams}, {Berger}  \&
  {Zauderer}}{{Williams} et~al.}{2013}]{Williams2013}
{Williams} P.~K.~G.,  {Berger} E.,   {Zauderer} B.~A.,  2013, \mn@doi [\apjl]
  {10.1088/2041-8205/767/2/L30}, \href
  {http://adsabs.harvard.edu/abs/2013ApJ...767L..30W} {767, L30}

\bibitem[\protect\citeauthoryear{{Williams}, {Gizis}  \& {Berger}}{{Williams}
  et~al.}{2017}]{Williams2017}
{Williams} P.~K.~G.,  {Gizis} J.~E.,   {Berger} E.,  2017, \mn@doi [\apj]
  {10.3847/1538-4357/834/2/117}, \href
  {http://adsabs.harvard.edu/abs/2017ApJ...834..117W} {834, 117}

\bibitem[\protect\citeauthoryear{{Winterhalter} et~al.,}{{Winterhalter}
  et~al.}{2006}]{Winterhalter2006}
{Winterhalter} D.,  et~al., 2006, in {Rucker} H.~O.,  {Kurth} W.,   {Mann} G.,
  eds, Planetary Radio Emissions VI. p.~595

\bibitem[\protect\citeauthoryear{{Yadav} \& {Thorngren}}{{Yadav} \&
  {Thorngren}}{2017}]{Yadav2017}
{Yadav} R.~K.,  {Thorngren} D.~P.,  2017, \mn@doi [\apjl]
  {10.3847/2041-8213/aa93fd}, \href
  {http://adsabs.harvard.edu/abs/2017ApJ...849L..12Y} {849, L12}

\bibitem[\protect\citeauthoryear{Yelle}{Yelle}{2004}]{Yelle2004}
Yelle R.~V.,  2004, Icarus, 170, 167

\bibitem[\protect\citeauthoryear{{Yelle}, {Young}, {Vervack}, {Young},
  {Pfister}  \& {Sandel}}{{Yelle} et~al.}{1996}]{Yelle1996}
{Yelle} R.~V.,  {Young} L.~A.,  {Vervack} R.~J.,  {Young} R.,  {Pfister} L.,
  {Sandel} B.~R.,  1996, \mn@doi [\jgr] {10.1029/95JE03384}, \href
  {http://adsabs.harvard.edu/abs/1996JGR...101.2149Y} {101, 2149}

\bibitem[\protect\citeauthoryear{{Zarka}}{{Zarka}}{1998}]{Zarka1998}
{Zarka} P.,  1998, \mn@doi [\jgr] {10.1029/98JE01323}, \href
  {http://adsabs.harvard.edu/abs/1998JGR...10320159Z} {103, 20159}

\makeatother
\end{thebibliography}




%
%


\bsp	
\label{lastpage}
\end{document}